\documentclass[12pt]{article}

\usepackage{graphicx}
\usepackage{amsmath}
\usepackage{amssymb}
\usepackage{bbm}
\usepackage{hyperref}
\usepackage{natbib}
\usepackage{multirow}
\usepackage{subcaption}
\usepackage{tikz}
\usepackage{titling}
\usetikzlibrary{shapes.geometric,arrows}

\newtheorem{theorem}{Theorem}[section]
\newtheorem{lemma}[theorem]{Lemma}
\newtheorem{proposition}[theorem]{Proposition}

\newcommand{\qed}{\nobreak \ifvmode \relax \else
      \ifdim\lastskip<1.5em \hskip-\lastskip
      \hskip1.5em plus0em minus0.5em \fi \nobreak
      \vrule height0.75em width0.5em depth0.25em\fi}

\def\bs{\boldsymbol}

\def\spacingset#1{\renewcommand{\baselinestretch}%
{#1}\small\normalsize} \spacingset{1}

\begin{document}

\title{Structured Shrinkage Priors}
\author{Maryclare Griffin \\
    Department of Mathematics and Statistics, \\
    University of Massachusetts Amherst,\\ Amherst, MA 01003, USA \\
    and \\
    Peter D. Hoff \\
    Department of Statistical Science,\\
     Duke University, \\ Durham, NC 27710, USA
}
\maketitle
\bigskip
\begin{abstract} 
	In many regression settings the unknown coefficients may have some known structure, for instance they may be ordered in space or correspond to a vectorized matrix or tensor. At the same time, the unknown coefficients may be sparse, with many nearly or exactly equal to zero. However, many commonly used priors and corresponding penalties for coefficients do not encourage simultaneously structured and sparse estimates. In this paper we develop structured shrinkage priors that generalize multivariate normal, Laplace, exponential power and normal-gamma priors. These priors allow the regression coefficients to be correlated a priori without sacrificing elementwise sparsity or shrinkage. The primary challenges in working with these  structured shrinkage priors are computational, as the corresponding penalties are intractable integrals and the full conditional distributions that are needed to approximate the posterior mode or simulate from the posterior distribution may be non-standard. We overcome these issues using a flexible elliptical slice sampling procedure, and demonstrate that these priors can be used to introduce structure while preserving sparsity.
\end{abstract}

\noindent {\it Keywords: Bayesian lasso, Lasso, multivariate Laplace, multivariate normal-scale mixture, sparsity, shrinkage.}

\vfill

\newpage 
\spacingset{1.45}

\section{Introduction}\label{sec:intro}

Shrinkage prior-based penalized estimates of regression coefficients are ubiquitous and useful.
When we observe an $n\times 1$ vector of responses $\bs y$ and an $n\times p$ matrix of regressors $\bs X$ and the data are high dimensional, i.e.\ $p$ is large relative $n$, traditional regression methods can fail. They may produce estimates of the $p\times 1$ vector of regression coefficients $\bs \beta$ have prohibitively large variance or are not unique because the data provide relatively little information about the unknown regression coefficients.

Using a prior for $\bs \beta$ that reflects our \emph{a priori} knowledge, we can obtain better estimates of $\bs \beta$.
When our \emph{a priori} knowledge involves similarities and differences among elements of $\bs \beta$, we might assume a structured mean-zero multivariate normal prior with covariance matrix $\bs \Sigma$. 
Alternatively, when our \emph{a priori} knowledge involves magnitudes of elements of $\bs \beta$, we might assume a sparsity inducing mean-zero independent Laplace prior with variance $\sigma^2$ in order to encode information about the origin and tail behavior of $\bs \beta$. This is useful when $\bs \beta$ is expected to be sparse, as the posterior mode of $\bs \beta$ under this prior corresponds to the $\ell_1$ or lasso penalized estimate of $\bs \beta$ \citep{Tibshirani1996,Park2008}.

When our \emph{a priori} knowledge involves both similarities and differences among and magnitudes of elements of $\bs \beta$, it would be desirable to assume a structured sparsity inducing shrinkage prior.
As an example, consider the analysis of brain-computer interface data.  
Brain-computer interfaces (BCIs) are used to detect changes in subjects' cognitive state from contemporaneous 
electroencephalography (EEG) measurements \color{black} at different channels corresponding to physical locations on the subject's skull\color{black}, which can be collected non-invasively at high temporal resolution \citep{Makeig2012,Wolpaw2012}.
We consider the P300 speller, a specific BCI device which is designed to detect when a subject is viewing a specified target letter \citep{Forney2013}.
For an individual subject, we consider $20$ indicators \color{black}$\bs y$ \color{black} for whether or not the subject was viewing a specified target letter during trial $i$ and $20$ contemporaneous EEG measurements \color{black}$\bs x_{jk}$ \color{black} from time point $j$ and channel $k$, for $j = 1,\dots, 208$ time points and $k = 1,\dots, 8$ channels.  A total of $240$ indicators and $240$ contemporaneous EEG measurements are available, but we consider a subset of $20$ indicators and contemporaneous EEG measurements because performance with such little data is of special interest and especially challenging. 
Scientifically, we expect to observe a P300 wave \color{black} in contemporaneous EEG measurements \color{black} during trials when the target letter is present, which is characterized by a sharp rise and then dip before returning to equilibrium. We expect that the wave will begin shortly after the target letter is shown, and will be observed earlier and more clearly on some channels than others. 

\begin{figure}[ht]
\color{black}
\centering
\includegraphics{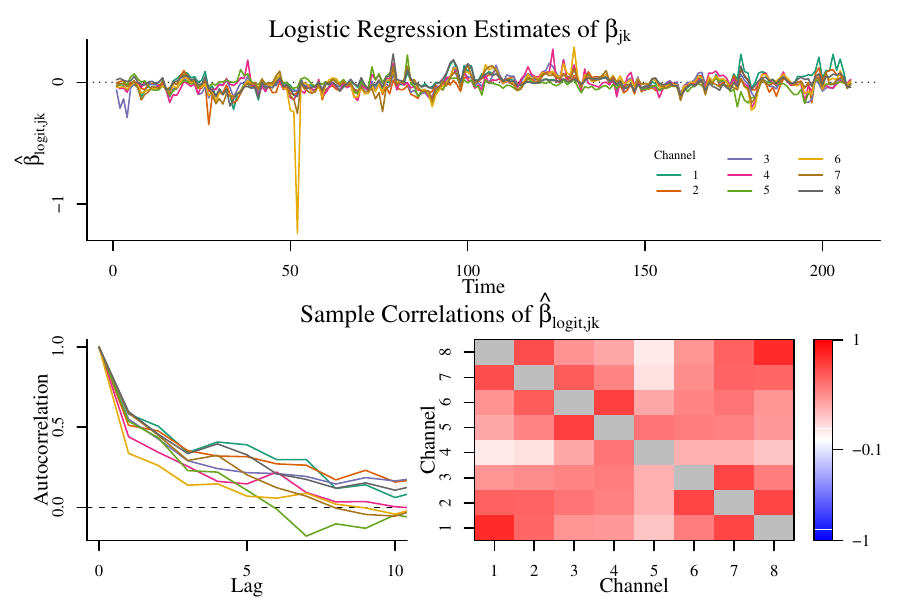} 
\caption{\color{black}Estimated logistic regression coefficients $\hat{\beta}_{logit,jk}$ using the first $20$ trials and their sample autocorrelations over time and sample correlations across channels.}
\label{fig:descc}
\end{figure}

Estimated regression coefficients that describe the relationship between EEG measurements from time point $j$ and channel $k$ and whether or not the subject was viewing the target letter can be obtained by regressing the indicators \color{black}$\bs y$ \color{black} on \color{black} an intercept and \color{black} measurements for each time point and channel \color{black}$\bs x_{jk}$ \color{black} using \color{black} a \color{black} logistic regression \color{black} model, which assumes $y_i \stackrel{indep.}{\sim} \text{Binomial} ((1 + \text{exp}\{-\gamma_{jk} - \beta_{jk} x_{ijk} \})^{-1})$\color{black}. 
Because EEG data are noisy\color{black}, making the P300 wave difficult to observe\color{black}, regression coefficients \color{black}$\beta_{jk}$ may be \color{black} poorly estimated. 
\color{black}I\color{black}t can be desirable to assume a model for the regression coefficients \color{black} that incorporates the scientific context suggesting that $\bs \beta$ should be sparse and structured, as only \color{black}$\beta_{jk}$ \color{black} that correspond to time points where the P300 wave occurs are expected to be nonzero and pairs of coefficients \color{black}$(\beta_{jk},\beta_{j'k'})$ \color{black} corresponding to \color{black} nearby \color{black} time points or channels are expected to be similar. \color{black} We observe evidence of such behavior in Figure~\ref{fig:descc}, which shows estimated logistic regression coefficients $\hat{\beta}_{logit,jk}$. \color{black}Consecutive channels correspond to neighboring locations; the appendix includes a map of channel locations. \color{black} Estimated regression coefficients corresponding to consecutive time points or neighboring channels tend to share the same sign. \color{black}

\color{black}This suggests a need for \color{black} structured shrinkage priors that (i) can incorporate \emph{a priori} knowledge of structure by allowing elements of $\bs \beta$ to covary with nondiagonal prior covariance matrix $\bs \Sigma = \mathbb{V}[\bs \beta]$, (ii) can incorporate \emph{a priori} knowledge of sparsity by allowing elementwise shrinkage of $\bs \beta$, and (iii) span the range of common structured and sparsity inducing priors by generalizing multivariate normal \color{black}and \color{black} independent Laplace priors.
However, existing approaches fail to satisfy all three of these criteria.  We can see this by considering the normal scale-mixture representations of many common priors, which are used in Bayesian approaches to sparse regression \citep{Polson2010}.
Letting `$\circ$' be the Hadamard elementwise product, $\bs z\sim \text{normal}(\bs 0, \bs \Omega)$ and $\bs s$ be a vector of stochastic scales that are independent of $\bs z$,  a prior distribution for $\bs \beta$ has a normal scale-mixture representation if there exists a density $p(\bs s | \bs \theta)$ such that $\bs \beta \stackrel{d}{=} \bs s \circ \bs z$. 
These priors are \color{black} interpretable \color{black} from a data generating perspective and have easy-to-compute moments. Specifically the prior covariance matrix of $\bs \beta$ that encodes the \emph{a priori} knowledge of structure is $\bs \Sigma = \mathbb{E}[\bs s \bs s']\circ \bs \Omega$. 

Much literature focuses on the unstructured case where $\bs s$ is a vector of independent, identically distributed elements and $\bs \Omega \propto \bs I_p$. This includes the Laplace prior, bridge/exponential power priors, normal-gamma priors, Dirichlet-Laplace priors, and horseshoe priors \citep{Park2008, Griffin2010, Carvalho2010, Bhattacharya2015}. 
These priors can incorporate \emph{a priori} knowledge of sparsity by modeling a separate stochastic scale for every element of $\bs \beta$ and choosing distributions for $\bs s$ that yield possibly sparse posterior modes. However, they do not allow elements of $\bs \beta$ to covary. Because the marginal covariance matrix $\bs \Sigma$ is an elementwise product of $\mathbb{E}[\bs s \bs s']$ and $\bs \Omega$, elements of $\bs \beta$ are uncorrelated \emph{a priori} under these priors.
The same limitation afflicts structured shrinkage priors which model elements of $\bs s$ as correlated but continue to assume that $\bs \Omega \propto \bs I_p$ \citep{vanGerven2010, Kalli2014, Wu2014, Zhao2016, Kowal2017}. 

A different strand of literature \color{black} includes all elliptically contoured prior distributions. It sets \color{black}  $\bs s = s \bs 1_p$ and models elements of $\bs z$ as \color{black} potentially \color{black} correlated with covariance matrix $\bs \Omega$.
\color{black}When $s^2$ is assumed to be exponentially distributed and $\bs \Omega \propto \bs I_p$, the prior corresponding to the group lasso introduced in \cite{Yuan2006} is obtained. The more general multivariate Laplace prior introduced by \cite{vanGerven2009} is obtained by assuming exponentially distributed $s^2$ and allowing arbitrary $\bs \Omega$. \color{black} These priors can incorporate \emph{a priori} knowledge of structure \color{black} in two ways: by treating elements of $\bs \beta$ as grouped through their shared stochastic scale $s$ and, when $\bs \Omega$ is allowed to take on arbitrary values, by encoding \emph{a priori} knowledge of structure through specification of $\bs \Omega$ that does not satisfy $\bs \Omega \propto \bs I_p$\color{black}, as $\bs \Sigma \propto \bs \Omega$. However, as these priors only have a single shrinkage factor $s$, they shrink all elements of $\bs \beta$ jointly and posterior modes based on these priors will only produce sparse estimates of $\bs \beta$ for which \color{black} $\bs \beta = \bs 0$ \citep{Simon2013}\color{black}.
Additionally, these priors do not generalize their independent counterparts. For instance, the multivariate Laplace prior with $\bs \Omega = \omega^2 \bs I_p$ does \emph{not} correspond to the independent Laplace prior, and the corresponding penalty is the group lasso as opposed to the lasso penalty. \color{black}This is not \color{black} appropriate \color{black} when \color{black} only \emph{some} elements of $\bs \beta$ are \color{black} expected to be \color{black} sparse.

At least one set of priors can incorporate \emph{a priori} knowledge of structure and sparsity by \color{black} also \color{black} modeling elements of the inverse covariance matrix of $\bs \Omega$.  This includes the prior distribution that yields the fused lasso penalty $\lambda_1 ||\bs \beta||_1 + \lambda_2 \sum_{i - j = 1}|\beta_j - \beta_i|$  and priors that correspond to more general structured penalties of the form $\bs \beta'\bs Q^{-1}\bs \beta + \lambda ||\bs \beta||_1$ where $\bs Q^{-1}$ is positive semidefinite \citep{Kyung2010, Ng2011, DeBrecht2012}.
The penalties corresponding to these priors are very popular, as they yield computationally feasible posterior mode optimization problems. Their main limitation is that relating the prior parameters $\lambda_1$ and $\lambda_2$ or $\bs Q^{-1}$ and $\lambda$ to the prior moments of $\bs \beta$ is prohibitively challenging. This makes \color{black} understanding \color{black} exactly how flexible these priors are and specifying values or priors for these parameters difficult, as it is unclear how to relate the kind of \emph{a priori} knowledge we might have to \color{black} the prior parameter values\color{black}.

One last class of relevant priors are the structured normal-gamma priors introduced in \cite{Griffin2012} and \cite{Griffin2012a}, which are obtained by assuming $\bs \beta \stackrel{d}{=} \bs C \left(\bs s \circ \bs z\right)$, where $\bs C$ is a $p\times q$ rectangular matrix with $p < q$ and $s^2_j$ and $z_j$ are independent gamma and normal random variables, respectively. Both elementwise and structured shrinkage can be simultaneously encouraged by setting $\bs C = \left[\bs I_p, \bs D\right]$, where $\bs D$ is a $p \times \left(q - p\right)$ matrix  with columns that encode groups of elements of $\bs \beta$ that should be penalized jointly. However, the theory that justifies the use of these structured priors is specific to generalizing independent normal-gamma priors. 

%
%

\color{black}W\color{black}e  construct a class of ``structured Hadamard product'' (SHP) priors that can incorporate \emph{a priori} knowledge of structure and sparsity by allowing \color{black} non-diagonal \color{black} $\bs \Omega$ and\color{black}, \color{black} accordingly\color{black}, \color{black} $\bs \Sigma$. 
The \color{black} main \color{black} challenge in using such priors is computational; \color{black} their development has been limited as a result\color{black}. The marginal prior\color{black}s \color{black} for $\bs \beta$ are intractable integrals and correspond to penalties $-\text{log}(\int p(\bs \beta | \bs \Omega, \bs s) p\left(\bs s | \bs \theta \right)d\bs s )$ with no simple closed form. \color{black} Their use \color{black}   requires computationally demanding Markov Chain Monte Carlo (MCMC) algorithms. Two exceptions are \cite{Finegold2011}, which develops a multivariate $t$-distribution by assuming \color{black}independent inverse-gamma \color{black} squared scales $s^2_j$, and \cite{Roy2021}, which develops \color{black} the structured product normal (SPN) prior \color{black} by assuming \color{black} normal \color{black} scales $\bs s\color{black}\sim \text{normal}\left(\bs 0, \bs \Psi\right)$.

This paper proceeds as follows.
In Section~\ref{sec:pri}, we describe the \color{black} SPN prior \color{black} and introduce \color{black}the novel \color{black} structured normal-gamma (SNG) and structured power/bridge (SPB) \color{black}priors which generalize the independent normal-gamma priors in \cite{Caron2008} and \cite{Griffin2010} and power/bridge priors in \cite{Frank1993} and \cite{Polson2014}, respectively.
\color{black}
We do not consider a structured generalization of the horseshoe prior because the corresponding prior covariance matrix $\bs \Sigma = \mathbb{E}[\bs s \bs s']\circ \bs \Omega$ is not finite, which makes it difficult to understand how \emph{a priori} knowledge of structure is incorporated. 
\color{black}
We discuss properties of these priors in Section~\ref{sec:pripro}. 
\color{black} Several properties and how they differ across the SPN, SNG, and SPB priors are previewed in Table~\ref{tab:properties}. \color{black}
In Section~\ref{sec:comp}, we describe how \color{black} the elliptical slice sampling  method of \cite{Murray2010} \color{black} can be used to overcome computational issues, regardless of the distribution assumed for elements of $\bs s$, and discuss estimation of hyperparameters.
\color{black}W\color{black}e focus on problems where the log-likelihood of the data given $\bs \beta$ and any additional nuisance parameters $\bs \phi$, denoted by $-h(\bs y | \bs X, \bs \beta, \bs \phi)$, can be written as conditionally quadratic in $\bs \beta$, i.e.\ $\text{exp}\{-h(\bs y | \bs X, \bs \beta, \bs \phi)\} \propto_{\bs \beta} \text{exp}\{-\frac{1}{2}(\bs \beta'\bs A \bs \beta - 2\bs \beta'\bs c)\}$ for some positive definite matrix $\bs A$ and real valued vector $\bs c$. This includes linear regression models \color{black} and \color{black} certain latent variable representations of logistic and negative binomial regression models \citep{Polson2013}. 
In Section~\ref{sec:num}, we use SHP priors to analyze the data depicted in Figure~\ref{fig:descc}.
A discussion follows in Section~\ref{sec:conc}.

\begin{table}
\footnotesize
\centering \color{black}
\begin{tabular}{cccc}
\textbf{Property} & \textbf{SNG} & \textbf{SPB} & \textbf{SPN} \\ \hline \hline
Generalizes an Independent Shrinkage Prior & \checkmark & \checkmark & \checkmark \\
Univariate Marginal is an Independent Shrinkage Prior & \checkmark & \checkmark & \checkmark \\
Generalizes a Laplace Prior & $c = 1$ &  $q = 1$ & \\
Generalizes a Normal Prior & $c \rightarrow \infty$ & $q\rightarrow 2$ &  \\
Infinite Spike or Pole at Zero & & $c \leq 1/2$ & \checkmark\\
Quadratic Scale Log Full Conditional & & & \checkmark \\
Arbitrary Covariance Structure Achievable & & & \checkmark \\ \hline
\end{tabular}
\caption{\color{black}Properties of specific SHP priors.}
\label{tab:properties}
\end{table}

\section{Structured Shrinkage Priors}\label{sec:pri}

We define several ``structured Hadamard product'' (SHP) prior distributions for $\bs \beta$ of the form $\bs \beta = \bs s \circ \bs z$, where `$\circ$' is the elementwise Hadamard product and $\bs z\sim \text{normal}(\bs 0, \bs \Omega)$ and $\bs s$ is a vector of stochastic scales that are independent of $\bs z$. Because the scales of elements of $\bs s$ are not separately identifiable from diagonal elements of $\bs \Omega$, we parametrize $\bs s$ such that $\mathbb{E}[s^2_j] = 1$ for $j = 1, \dots, p$. These priors are mean $\bs 0$ and have prior variance $\bs \Sigma = \mathbb{E}[\bs s \bs s'] \circ \bs \Omega$.

\paragraph{Structured Product Normal (SPN) Prior:}

When $\bs s \sim \text{normal}(\bs 0, \bs \Psi)$ and $\bs \Psi$ is a positive definite matrix with diagonal elements equal to $1$, the SPN prior is obtained. \color{black}T\color{black}he parameters of the prior distribution for $\bs s$, denoted by $\bs \theta$, correspond to the off-diagonal elements of $\bs \Psi$.
\color{black}E\color{black}lements of the hyperparameters $\bs \Omega$ and $\bs \Psi$ can be related to prior moments of $\bs \beta$. Diagonal elements of $\bs \Omega$ correspond to prior variances of $\bs \beta$, and off diagonal elements of $\bs \Omega$ and $\bs \Psi$ determine covariances and fourth-order prior cross moments of $\bs \beta$, as shown in the appendix. Originally discussed as a sparsity inducing penalty in \cite{Hoff2016b} and later implemented as a prior distribution in \cite{Roy2021}, the SPN prior is uniquely computationally simple to work with as the full conditional distributions of $\bs z$ and $\bs s$ are both multivariate normal distributions when the log-likelihood  $-h(\bs y | \bs X, \bs \beta, \bs \phi )$ is quadratic or conditionally quadratic in $\bs \beta$. This is described in greater detail in Section~\ref{sec:comp}.
The SPN prior is also appealing as it is the only prior we define that has correlations among elements of $\bs s$. To reduce the number of freely varying parameters and to facilitate use of the SPN prior in settings where it is challenging to estimate or formulate prior opinions about fourth-order prior cross moments of $\bs \beta$, we also define a special case of the SPN prior which \color{black} we call the symmetric SPN (sSPN) prior. The sSPN prior \color{black} requires that all elements of $\bs \Psi$ be positive and that the correlation matrix corresponding to $\bs \Omega$ has elements that are equal in magnitude to the elements of $\bs \Psi$, i.e.\ $|\omega_{ij}/\sqrt{\omega_{ii}\omega_{jj}}| = \psi_{ij}$. Under this constraint, $\bs \Psi$ is a deterministic function of $\bs \Omega$ \color{black}and \color{black} $\bs \theta$ is an empty vector.


\paragraph{Structured Normal-Gamma (SNG) Prior:} 

When the \color{black} squared \color{black} stochastic scales $s^2_j$ are independent gamma random variables $s^2_j \sim \text{gamma}(c, c)$ with $\mathbb{E}[s^2_j] = 1$ for fixed $c\in (0, \infty)$ \color{black} and the stochastic scales $s_j = \sqrt{s^2_j}$ are strictly positive\color{black}, the structured normal-gamma (SNG) prior is obtained. The SNG prior \color{black} generalizes \color{black} the normal-gamma prior of \cite{Griffin2010}, \color{black}which \color{black} is obtained by setting $\bs \Omega \propto \bs I_p$. The \color{black} fixed \color{black} shape parameter $c$  parametrizes the prior class \color{black} and \color{black} $\bs \theta$ is an empty vector. The value chosen for $c$ determines the prior fourth order moments of $\bs \beta$; SNG priors with smaller values of $c$ have lighter tails.

\paragraph{Structured Power/Bridge (SPB) Prior:} 
When the \color{black} squared \color{black} stochastic scales $s^{2}_j$ are independently distributed according to a polynomially tilted positive $\alpha$-stable distribution with index of stability $\alpha = q/2$ and $\mathbb{E}[s^2_j] = 1$ for fixed $q\in(0, 2)$ \color{black} and the stochastic scales $s_j = \sqrt{s^2_j}$ are strictly positive\color{black}, the structured power/bridge (SPB) prior is obtained.
The SPB prior generalizes the bridge or exponential power prior discussed in \cite{Polson2014}. The \color{black} fixed \color{black} shape parameter $q$ parametrizes the prior class \color{black} and \color{black}  $\bs \theta$ is an empty vector. Working with this prior is especially computationally challenging because the polynomially tilted positive $\alpha$-stable density $p(s_j | \bs \theta)$ is not available in closed form.  Fortunately, a polynomially tilted positive $\alpha$-stable \color{black} random variable \color{black} can be represented as a rate mixture of generalized gamma random variables  \citep{Devroye2009}. A more detailed description of this representation and how it enables computation under the SPB prior is provided in the appendix.
	As with the SNG prior, the value chosen for $q$ for the SPB prior determines the prior fourth order moments of $\bs \beta$; SPB priors with larger values of $q$ have lighter tails.


\begin{figure}[ht]
\begin{center}
\begin{tikzpicture}
\tikzstyle{old}=[draw,rectangle, rounded corners, fill=gray!20,
minimum width=3cm, minimum height=1cm]
\tikzstyle{new}=[draw, rectangle, rounded corners, fill=white, minimum width=2cm, aspect=2, minimum height=1cm]
\tikzstyle{a}=[thick]
\node[old] (laplace) at (-5, -2) {IID Laplace};
\node[old] (normal) at (5, -1) {Multivariate Normal};
\node[new] (sng) at (0, -4) {SNG};
\node[new] (spb) at (0, 0) {SPB};
\node[new] (spn) at (7, -4) {SPN};
\draw[a] (sng) -- node[midway,above,sloped] {
\color{black}$c\rightarrow \infty$
} (normal);
\draw[a] (spb) -- node[midway,above,sloped] {
\color{black}$q\rightarrow 2$
} (normal);
\draw[a] (sng) -- node[midway,above,sloped] {
\color{black}$c = q = 1$
} (spb);
\draw[a] (sng) -- node[midway,above,sloped] {
\color{black}$c=1/2$, $\bs \Omega$ and $\bs \Psi$ diagonal
} (spn);
\draw[a] (sng) -- node[midway,above,sloped] {
\color{black}$c=1$, $\bs \Omega \propto \bs I_p$
} (laplace);
\draw[a] (spb) -- node[midway,above,sloped] {
\color{black}$q=1$, $\bs \Omega \propto \bs I_p$
} (laplace);
\end{tikzpicture}
\end{center}
\caption{Relationships between structured product normal (SPN), structured normal-gamma (SNG) and structured power/bridge (SPB) shrinkage priors and independent, identically distributed (IID) Laplace and multivariate normal priors.}
\label{fig:chart}
\end{figure}

\paragraph{Relationships Between Priors:} 
\color{black}Figure~\ref{fig:chart} shows relationships between the SPN, SNG, and SPB priors, the independent Laplace \color{black} prior, \color{black} and \color{black} the \color{black} multivariate normal \color{black} prior\color{black}. 
\color{black}When $c = q = 1$, t\color{black}he SNG and SPB priors are equivalent and generalize the independent Laplace prior. \color{black}In the limit as $c \rightarrow \infty$ or $q \rightarrow 2$, \color{black} the SNG and SPB priors generalize the multivariate normal prior. The SPN prior does not generalize the Laplace prior or the multivariate normal prior, but is equivalent to the SNG prior with $c = 1/2$ when $\bs \Psi$ and $\bs \Omega$ are diagonal.

\section{Properties}\label{sec:pripro}

\subsection{Univariate Prior Properties}

\color{black}I\color{black}ntroducing structure does not alter the marginal prior distributions of elements $\beta_j$.
For instance, if we assume \color{black}a structured prior that generalizes the independent Laplace prior for \color{black} $\bs \beta$, \color{black}e.g. an \color{black} SNG prior with $c = 1$ or an SPB prior with $q = 1$, the marginal prior distribution of any element $\beta_j$ is Laplace. This \color{black}follows \color{black} from the stochastic representation of elements $\bs \beta$ under these priors, $\bs \beta = \bs s \circ \bs z$. Recall that $\bs s$ is a vector of stochastic scales, $\bs z \sim \text{normal}(\bs 0, \bs \Omega)$, $\bs s$ and $\bs z$ are independent of each other, and that all three SHP priors are obtained by assuming different distributions for $\bs s$. 
If we consider an individual element $\beta_j = s_j z_j$ under the SNG or SPB priors, introducing structure does not affect $s_j$ at all and does not affect the marginal distribution of $z_j$, as the marginal distribution of an element of a correlated normal vector is still normal.
Similarly, if we consider an individual element $\beta_j = s_j z_j$ under the SPN prior, introducing structure does not affect the marginal distributions of $s_j$ and $z_j$, again because the marginal distribution of an element of a correlated normal vector is still normal.

Because introducing structure does not alter the marginal prior distributions of elements $\beta_j$, the marginal prior distributions of elements $\beta_j$ retain the same sparsity inducing properties of the corresponding independent priors.
\color{black}For instance, \color{black}independent normal-gamma priors with $c \leq 1/2$ are known to have an infinite spike or pole at $b_j = 0$, which has been viewed in the literature as a sufficient condition for the recovery of sparse signals \citep{Carvalho2010, Griffin2010, Bhattacharya2015}. 
Because introducing structure does not alter the marginal prior distributions of elements $\beta_j$, the marginal prior distributions of elements $\beta_j$ under SNG priors with $c \leq 1/2$ and under the SPN prior will also have an infinite spike or pole at $b_j = 0$. \color{black}Proofs are provided \color{black} in the appendix.

\subsection{Joint Prior Properties}

\subsubsection{Range of Achievable $\bs \Sigma$}

\color{black}I\color{black}ntroducing structure while retaining elementwise shrinkage can come at a cost. 
Specifically, under the SNG, SPB and SPN priors, preserving elementwise shrinkage limits how correlated elements of $\bs \beta$ can be. Recall that under all three SHP priors, the prior covariance matrix of $\bs \beta$ is $\bs \Sigma = \mathbb{E}[\bs s \bs s'] \circ \bs \Omega$. 
Under the SNG and SPB priors, $\mathbb{E}[\bs s \bs s']$ is constant for fixed values of $c$ or $q$.
Diagonal elements of $\mathbb{E}[\bs s \bs s']$ are equal to $1$, whereas off-diagonal elements are less than one in absolute value. Thus, $\mathbb{E}[\bs s \bs s']$ shrinks off-diagonal elements of $\bs \Omega$, reducing dependence. When $\bs \beta \in \mathbb{R}^2$, we can explicitly calculate the maximum marginal prior correlation $\rho$ under the SNG and SPB priors as a function of $c$ or $q$. Under the SNG prior, the maximum correlation is equal to $c^{-1}(\Gamma(c + 1/2)/\Gamma(c))^2$ and under the SPB prior, the maximum correlation is equal to $(\pi/2)(\Gamma(2/q)/\sqrt{\Gamma(1/q)\Gamma(3/q)})^2$. When $q = c = 1$ and both priors are equivalent, the maximum correlation is equal to $\pi/4 \approx 0.79$. 

\begin{figure}
\centering
\includegraphics{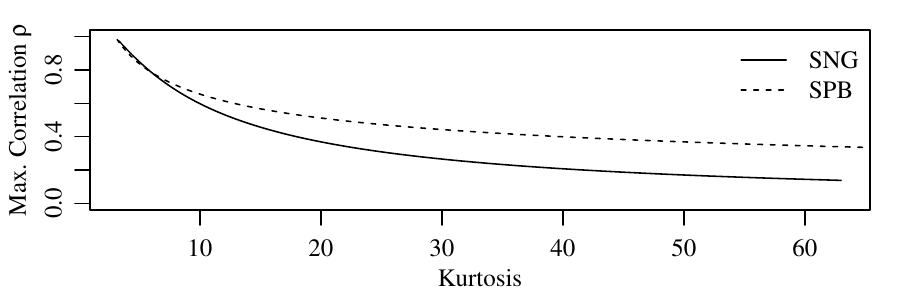}
\caption{Maximum marginal prior correlation $\rho$ for $\bs \beta \in \mathbb{R}^2$ as a function of kurtosis.}
\label{fig:maxcorr}
\end{figure}

We plot the maximum correlation as a function of kurtosis, a measure of tail behavior, under both priors in Figure~\ref{fig:maxcorr}. We observe greater reductions in the maximum correlation when the kurtosis is higher, and under the SNG prior relative to a SPB prior with equal kurtosis. The restricted range of $\bs \Sigma$ under the SNG and SPB priors is similar to the restricted range of the variance-covariance matrix of the alternative multivariate $t$-distribution introduced in \cite{Finegold2011}. Intuitively, the restricted range of $\bs \Sigma$ relates to the conflict that arises between the properties of the marginal joint density $p(\bs \beta | \bs \Omega, \bs \theta )$ needed to preserve elementwise  shrinkage, specifically concentration of the density along the axes, and the properties of the marginal joint density needed to encourage structure, specifically concentration of the density along a line when $p = 2$.

In contrast, the unrestricted SPN prior can accommodate an arbitrary prior covariance $\bs \Sigma$.
Given a positive semidefinite prior covariance $\bs \Sigma$, it is always possible to find at least one pair of positive semidefinite covariance matrices $\bs \Omega$ and $\bs \Psi$ that satisfy $\bs \Sigma = \bs \Omega \circ \bs \Psi$ \citep{Styan1973}.
The sSPN prior is less flexible. It is easy to simulate a positive semidefinite covariance matrix $\bs \Sigma$ for which the corresponding values of $\bs \Omega$ or $\bs \Psi$ satisfying $| \omega_{ij}| =| \psi_{ij}|$ are not positive semidefinite, but challenging to explicitly characterize the class of covariance matrices $\bs \Sigma$ that correspond to non-positive semidefinite values of $\bs \Omega$ or $\bs \Psi$.

\subsubsection{Copulas}

\begin{figure}
\centering
\vspace{-5mm} 
\includegraphics{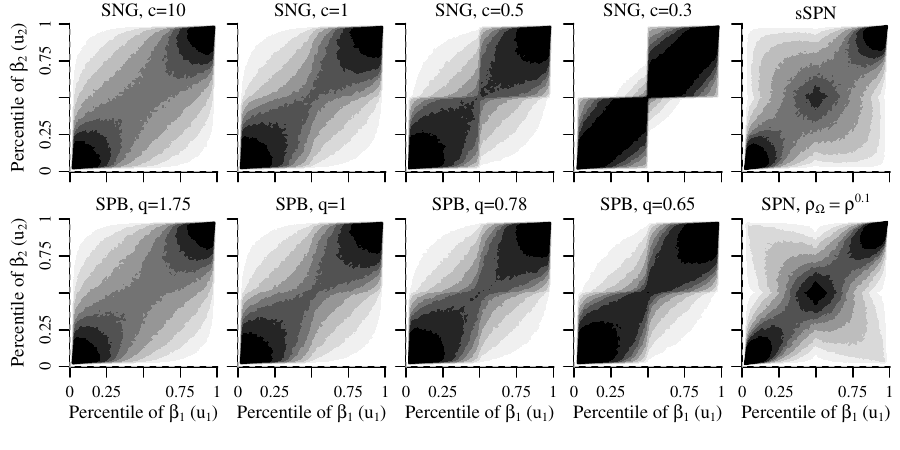}
\caption{Approximate copula densities for $\bs \beta \in \mathbb{R}^2$ with unit marginal prior variances and marginal prior correlation $\rho = 0.5$. Copula density approximations were made by simulating $1,000,000$ values from the corresponding prior, transforming simulated values of $\beta_1$ and $\beta_2$ into percentiles $u_1$ and $u_2$, and computing a kernel bivariate density estimate of the percentiles. 
}
\label{fig:cop}
\end{figure}

Even when all three SHP priors share the same prior covariance matrix $\bs \Sigma$, the induced dependence structures vary widely.
We compare the induced dependence structures by considering $\bs \beta \in \mathbb{R}^2$ with unit marginal variances and correlation $\rho = 0.5$, and examining corresponding copula densities. Let $F^{SNG,1}_{j}(\beta_j)$ refer to the marginal prior CDF of $\beta_j$ corresponding to one of the SHP prior distributions. We can always write $\beta_j \stackrel{d}{=}F^{-1}_j(u_j)$, where $u_j$ is a random variable with uniform margins. The joint distribution of the $p\times 1$ vector $\bs u$ is called the copula of $\bs \beta$, and it characterizes the induced dependence structure.
Even if the inverse CDF's $F^{-1}_j(u_j)$ are not known, the copula density can be approximated by simulating values of $\bs \beta$, transforming simulated values of $\beta_1$ and $\beta_2$ into percentiles $u_1$ and $u_2$, and computing a kernel bivariate density estimate of the percentiles.
Figure~\ref{fig:cop} shows numerical approximations to copula densities for several SHP priors with unit marginal prior variances and marginal prior correlation $\rho = 0.5$.

The first four panels in the top and bottom rows show copula densities under SNG and SPB priors with increasingly heavy tails, and the final panels on the top and bottom show \color{black} copula densities \color{black} under SPN priors with $\color{black}\omega_{12}\color{black} = \rho^{0.5}$ and $\color{black}\omega_{12}\color{black}= \rho^{0.1}$. The parameters of the SPB priors have been chosen to ensure that the kurtosis, a measure of tail behavior, of each SPB prior is equal to the kurtosis of the SNG prior above it,\  e.g. the SNG prior with $c = 0.3$ has the same kurtosis as the SPB prior with $q = 0.65$.

The dependence structures induced by the SNG and SPB priors are similar. 
As $c$ or $q \rightarrow 0$ and the SNG and SPB priors become less normal and heavier tailed, the copulas display increasingly strong orthant dependence. 
This means that as $c$ or $q \rightarrow 0$, the priors concentrate more strongly around values of $\bs \beta$ that have the same sign.
The SNG prior displays especially strong orthant dependence and appears to converge to a uniform distribution over the positive and negative orthants as $c \rightarrow 0$.
Additionally, as $c$ or $q \rightarrow 0$ the copula densities become more concentrated around the axes where at least one element of $\bs \beta$ is nearly or exactly equal to zero, suggesting that these priors are still encouraging elementwise shrinkage.

Like the SNG and SPB priors, the SPN priors concentrate in the upper-right and lower-left corners, where $\beta_1$ and $\beta_2$ have the same sign and are large in magnitude. However in comparison to the SNG and SPB priors, both SPN priors also concentrate strongly at the origin, where $\beta_1 = \beta_2 = 0$, and less strongly along the axes, where $\beta_1 = 0$ or $\beta_2 = 0$. Both SPN priors also concentrate more about values of $\bs \beta$ that are similar in magnitude and opposite in sign. The extent of this depends on how $\color{black}\omega_{12}\color{black}$ is chosen. Compared to the sSPN prior which sets $\color{black}\omega_{12}\color{black}$ and $\color{black}\psi_{12}\color{black}$ symmetrically such that  $\color{black}\omega_{12}\color{black}=\color{black}\psi_{12}\color{black}$, the SPN prior with $\color{black}\omega_{12}\color{black} = \rho^{0.1}$ and $\color{black}\psi_{12}\color{black} = \rho^{0.9}$ concentrates more strongly about values of $\bs \beta$ that are equal in magnitude but opposite in sign.
Intuitively, this is due to the fact that $\bs \beta$ is made up of one strongly correlated component and one very weakly correlated component, both of which can take any value on $\mathbb{R}$ when $\color{black}\omega_{12}\color{black} = \rho^{0.1}$ and $\color{black}\psi_{12}\color{black} = \rho^{0.9}$.

\subsubsection{Conditional Prior Distributions}

\begin{figure}[ht]
\centering
\includegraphics{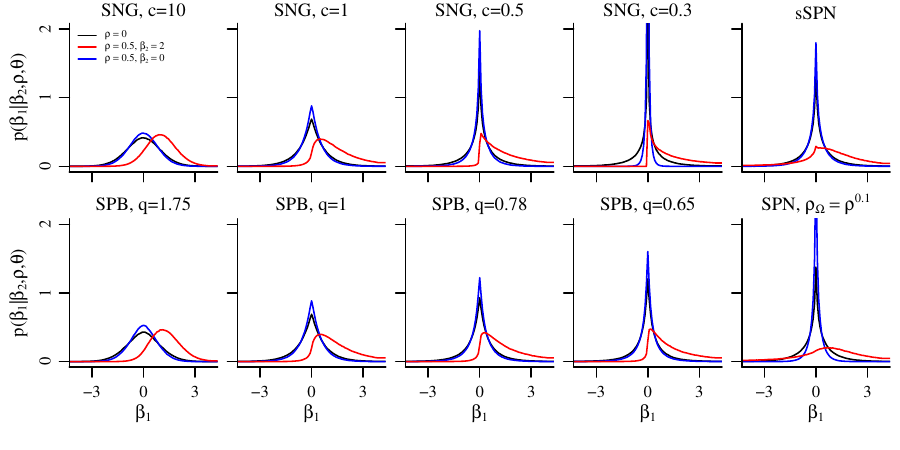}
\caption{Conditional prior distributions $p(\beta_1 | \beta_2, \rho, \bs \theta )$ for $\bs \beta \in \mathbb{R}^2$ with unit marginal prior variances, marginal prior correlation $\rho \in \{0, 0.5\}$ and $\beta_2 \in \{0, 2\}$ obtained via simulation.}
\label{fig:cond}
\end{figure}

We can explore how introducing structure can inform shrinkage of elements of $\bs \beta$ by examining the induced conditional prior distributions.
Figure~\ref{fig:cond} shows conditional prior distributions $p(\beta_1 | \beta_2, \bs \Omega, \bs \theta )$, for $\bs \beta \in \mathbb{R}^2$ with unit marginal prior variance, marginal prior correlations $\rho =\{0, 0.5\}$, and $\beta_2 = \{0, 2\}$.
Again, the first four panels in the top and bottom rows show conditional prior distributions under SNG and SPB priors with increasingly heavy tails, and the final panels on the top and bottom show conditional prior distributions under SPN priors with $\color{black}\omega_{12}\color{black} = \rho^{0.5}$ and $\color{black}\omega_{12}\color{black} = \rho^{0.1}$. The parameters of the SPB priors have been chosen to ensure that the kurtosis, a measure of tail behavior, of each SPB prior is equal to the kurtosis of the SNG prior above it,\  e.g. the SNG prior with $c = 0.3$ has the same kurtosis as the SPB prior with $q = 0.65$.

Introducing structure via a positive prior correlation between $\beta_1$ and $\beta_2$ can allow us to use knowledge of $\beta_2$ to better estimate $\beta_1$. We see that sparsity inducing SHP priors concentrate more strongly about zero than their independent counterparts when $\beta_2 = 0$, and shift their mass towards positive nonzero values of $\beta_1$ when $\beta_2 = 2$.
The conditional priors reflect differing amounts of origin versus tail dependence.
The nearly normal SNG and SPB priors on the far left display relatively more tail dependence, as the structured conditional priors for $\beta_1$ differ more from their independent counterparts when $\beta_2 = 2$ than when $\beta_2 = 0$.
At the other end of the spectrum, the heavier-than-Laplace tailed SNG and SPB priors and the sSPN prior display more origin dependence than tail dependence, as the structured conditional priors for $\beta_1$ concentrate much more strongly about $\beta_1 = 0$ when $\beta_2 = 0$ than their independent counterparts, and still have a \color{black} mode \color{black} near $\beta_1 = 0$ even when $\beta_2 = 2$.
This is especially striking under the SNG priors with $c \leq 1/2$ and the sSPN prior, which retain a very sharp peak at $\beta_1 = 0$ even when $\beta_2 = 2$.
The asymmetric SPN prior with $\color{black}\omega_{12}\color{black} = \rho^{0.1}$ is in between; the full conditional distribution of $\beta_1$ concentrates much more strongly about $\beta_1 = 0$ when $\beta_2=0$ and shifts markedly to towards $\beta_1=2$ when $\beta_2 = 2$.

\subsubsection{Joint Marginal Prior Contours}

\begin{figure}
\centering
\includegraphics{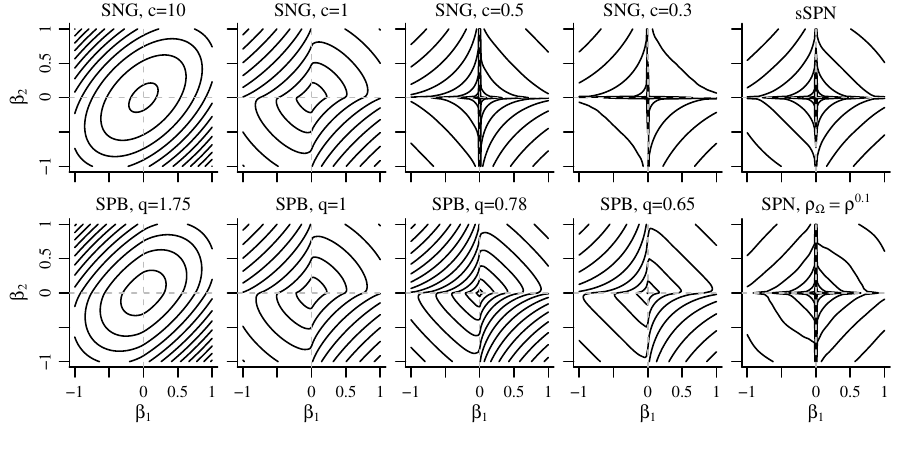}
\caption{Prior density contours for $\bs \beta \in \mathbb{R}^2$ with unit marginal prior variance and marginal prior correlation $\rho = 0.5$.}
\label{fig:pencon}
\end{figure}

The joint marginal prior contours, which can be interpreted as contours of a penalty function, provide a powerful tool for studying the properties of different priors.
Figure~\ref{fig:pencon} shows Monte Carlo approximations of the contours of the joint marginal priors $\int p(\bs \beta | \bs s, \bs \Omega)p(\bs s | \bs \theta)d\bs s$, for $\bs \beta \in \mathbb{R}^2$ with marginal prior variance-covariance matrix $\bs \Sigma = (1 - \rho)\bs I_2 + \rho\bs 1_2 \bs 1_2'$ and marginal prior correlation $\rho = 0.5$. 

Again, the first four panels in the top and bottom rows show contours of SNG and SPB priors with increasingly heavy tails. The parameters of the SPB priors have been chosen to ensure that the kurtosis, a measure of tail behavior, of each SPB prior is equal to the kurtosis of the SNG prior above it,\  e.g. the SNG prior with $c = 0.3$ has the same kurtosis as the SPB prior with $q = 0.65$.
The rightmost panels show contours of SPN priors. The top panel is a symmetric sSPN prior with $\color{black}\omega_{12}\color{black} = \rho^{0.5}$ and $\color{black}\psi_{12}\color{black} = \rho^{0.5}$, whereas the bottom panel is an asymmetric  SPN prior with $\color{black}\omega_{12}\color{black} = \rho^{0.1}$ and $\color{black}\psi_{12}\color{black} = \rho^{0.9}$.

All three SHP priors encourage similar estimates of $\beta_1$ and $\beta_2$ by pushing contours away from the origin when $\beta_1$ and $\beta_2$ have the same sign and pushing the contours towards the origin when $\beta_1$ and $\beta_2$ have opposite signs. The SPN, SNG with $c \leq 1$ and SPB with $q \leq 1$ priors encourage sparse estimates of $\bs \beta$ by retaining discontinuities of the log marginal prior on the axes.
The contours do not necessarily keep the same shape as the value of prior density changes. Contours closer to the origin are more similar to their independent counterparts, with relatively more encouragement of sparsity than structure, whereas contours farther from the origin tend to encourage relatively more structure and less sparsity. This is especially evident under the SPN prior with $\rho_{\omega} = \rho^{0.1}$. 
It is clear from the contours that these priors are \emph{not} log-concave when $\rho \neq 0$ and  correspond to \emph{non-convex} penalties \color{black} for \color{black} SNG priors with $c \leq 1$, SPB priors with $q \leq 1$, and SPN and sSPN priors. Under the SNG priors with $c \leq 1/2$ and the SPN priors, the contours do not cross the axes. 
The joint marginal distribution of $\bs \beta$ under the SNG prior with $c \leq 1/2$ has has an infinite spike or pole along the axes,  where at least one element is equal to $0$.

\begin{proposition}
For the SNG prior \color{black}with \color{black} $c \color{black}\leq\color{black} 1/2$ and \color{black}any \color{black} $b_j = 0$, then $p( \bs b | c, \bs \Omega) = +\infty$.
\end{proposition}

This is a multivariate analogue of the univariate marginal prior's spike or infinite pole at $b=0$ under the SNG prior with $c \leq 1/2$. The joint marginal SPN prior behaves similarly. 

\begin{proposition}
For the SPN prior, if \color{black} any \color{black} $b_j = 0$, then $p(\bs b | \bs \Psi, \bs \Omega) = +\infty$.
\end{proposition}
Proofs are given in the appendix. This suggests that the SPN and SNG priors with $c \color{black}\leq\color{black} 1/2$ may recover sparse signals well \citep{Carvalho2010, Griffin2010, Bhattacharya2015}. However the presence of an infinite spike or pole at $b_j = 0$ also makes the already challenging problem of computing the posterior mode intractable, as the log marginal prior is not only nonconvex but also has infinitely many modes.

\subsection{Posterior Properties}

One last perspective on the properties of all three SHP priors can be gained by examining how the posterior mode of $\bs \beta$ relates to the unpenalized OLS estimate $\hat{\beta}_{OLS}$. This can offer an intuitive understanding of the properties of estimators obtained under the three different priors. In the sparsity inducing penalty literature this is often characterized by the thresholding function, which is defined in the context of linear model for $\bs y$ with a single covariate $\bs x$ that satisfies $\bs x'\bs x = 1$ as
\begin{align}\label{eq:modethresh}
\hat{\beta}=\text{argmin}_{\beta} \frac{1}{2\phi^2}\left(\hat{\beta}_{OLS}-\beta\right)^2 \color{black}-\color{black} \text{log}\left(\int p\left(\beta | s, \omega \right)p\left(s | \theta\right) d s \right),
\end{align}
where $\phi^2$ refers to assumed variance of deviations of $\hat{\beta}_{OLS} - \beta$.
For example, \eqref{eq:modethresh} gives the soft-thresholding function when $\beta$ has a mean-zero Laplace distribution.

Because the advantage of working with any of the three three SHP priors is the ability to introduce a priori dependence across elements of vector valued $\bs \beta$, examining the corresponding univariate thresholding functions for scalar $\beta$ given by \eqref{eq:modethresh} does not fully explore how the posterior mode of $\bs \beta$ relates to the unpenalized OLS estimate $\hat{\bs \beta}$. Accordingly, we define and examine a bivariate thresholding function. We continue to assume a linear model for $\bs y$, but assume an orthogonal design matrix comprised of two covariates $\bs X$ that satisfies $\bs X'\bs X = \text{diag}\left\{\bs 1_2 \right\}$ instead of a single covariate $\bs x$. The bivariate thresholding function relates the posterior mode of both $\beta_1$ and $\beta_2$ to the noise variance $\phi^2$, prior parameters $\bs \Omega$ and $\bs \theta$, and OLS estimates $\hat{\beta}_{OLS,1}$ and $\hat{\beta}_{OLS,2}$, and is given by

\begin{align}\label{eq:sspmodethresh}
\hat{\bs \beta}=\text{argmin}_{\bs \beta} \frac{1}{2\phi^2}\left|\left|\hat{\bs \beta}_{OLS} - \bs \beta\right|\right|^2_2 \color{black}-\color{black} \text{log}\left(\int p\left(\bs \beta | \bs s, \bs \Omega \right)p\left(\bs s | \bs \theta\right) d \bs s \right).
\end{align}

\begin{figure}[ht]
\centering
\includegraphics{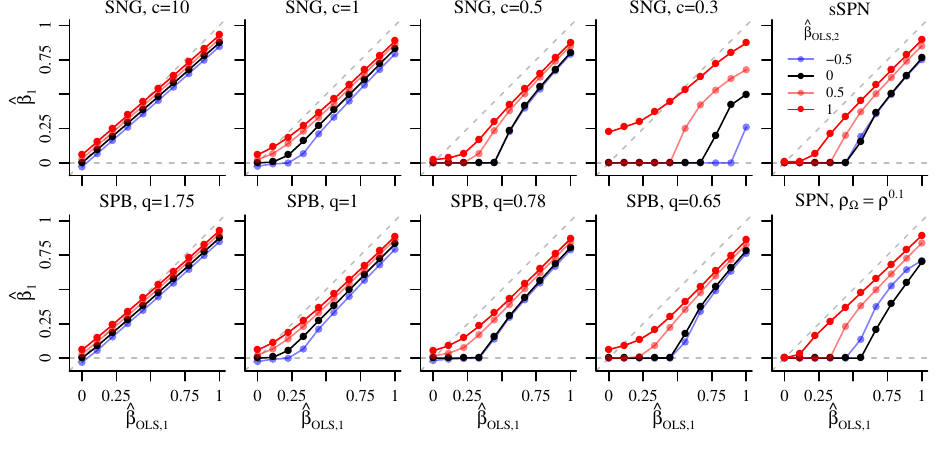}
\caption{Approximate bivariate thresholding functions computed according to \eqref{eq:sspmodethresh} for $\bs \beta \in \mathbb{R}^2$ with unit marginal prior variances, marginal prior correlation $\rho=0.5$, noise variance $\phi^2 = 0.1$, $\hat{\beta}_{OLS,2} \in \{-0.5, 0, 0.5, 1\}$ and $\hat{\beta}_{OLS,1} \in (0, 1)$.}
\label{fig:thresholding}
\end{figure}

\color{black}T\color{black}he bivariate thresholding function can be approximated using a Gibbs-within-EM algorithm described in the Section~\ref{sec:comp}.
Figure~\ref{fig:thresholding} shows approximate bivariate thresholding functions for $\bs \beta \in \mathbb{R}^2$ with unit marginal prior variances, marginal prior correlation $\rho=0.5$, noise variance $\phi^2 = 0.1$, $\hat{\beta}_{OLS,2} \in \{-0.5, 0, 0.5, 1\}$ and $\hat{\beta}_{OLS,1} \in \left[0, 1\right]$,  computed from $1,001,000$ Gibbs sampler iterations, with the first $1,000$ samples discarded as burn-in. 

Examination of the bivariate thresholding functions confirms that the SPN prior, SNG prior with $c \leq 1$ and SPB prior with $c \leq 1$ can yield possibly sparse posterior mode estimates of $\bs \beta$, and that the introduction of structure encourages or discourages sparse estimates $\hat{\beta}_1$ depending not only on the observed value of $\hat{\beta}_{OLS,1}$ but also the observed value of $\hat{\beta}_{OLS,2}$.
There are a few interesting trends that relate to previously identified properties of the SPN, SNG and SPB priors.
\color{black}B\color{black}oth SPN priors shrink $\hat{\beta}_1$ towards $0$ \color{black}less \color{black} aggressively when $\hat{\beta}_{OLS,2} = -0.5$ than when $\hat{\beta}_{OLS,2} = 0$, which reflects the tendency of the SPN priors to encourage estimates of $\bs \beta$ with similar \color{black}absolute \color{black} magnitudes.
Additionally, the value of $\hat{\beta}_{OLS,2}$ appears to affect the estimate $\hat{\beta}_1$ more when $\hat{\beta}_{OLS,1}$ is smaller under the SPN prior, SNG prior with $c \leq 1$ and SPB prior with $c \leq 1$. This is consistent with what we observed when examining the copulas and conditional prior distributions; as the tails become heavier, the Laplace and heavier-than Laplace prior distributions display more dependence at the origin than the tails. 
Last, the SNG prior with $c = 0.3$ not only induces especially strong shrinkage of $\hat{\beta}_1$ when $\hat{\beta}_{OLS,2}$ is small, but also induces aggressive inflation of $\hat{\beta}_1$ when $\hat{\beta}_{OLS,1}$ is small and $\hat{\beta}_{OLS,2}$ is very large. 
This makes sense in the context of the prior conditional distributions under the SNG and SPB priors with $c = 0.3$ and $q = 0.65$. Although these two priors have equally heavy tails, the SNG prior is much more concentrated at the origin and has heavier tails.

\section{Computation}\label{sec:comp}

\subsection{Posterior Approximation}

The posterior mode of $\bs \beta$ maximizes the integral $\int p(\bs y | \bs X,\bs \beta, \bs \phi) p(\bs \beta |\bs \Omega, \bs s) p(\bs s | \bs \theta ) d\bs s$ over $\bs \beta$ \color{black}and can be thought of as a \color{black} penalized estimate of $\bs \beta$, where the penalty is given by $-\text{log}(\int p(\bs \beta |\bs \Omega, \bs s) p(\bs s | \bs \theta ) d\bs s )$.
This integral is generally intractable when $\bs \Omega$ is not diagonal, but can be maximized using an MCMC within Expectation Maximization (EM) algorithm \citep{Dempster1977}.  Given an initial value $\bs \beta^{(0)}$, this algorithm proceeds by iterating the following until $||\bs \beta^{(i + 1)} - \bs \beta^{(i)}||^2_2$ converges:
\begin{itemize}
	\item using MCMC to simulate $M$ draws $\bs s^{(1)}, \dots, \bs s^{(M)}$ from the full conditional distribution of $\bs s$ given $\bs \beta^{(i)}$ and $\bs \theta$, set $\widehat{\mathbb{E}}[(\frac{\bs 1}{\bs s}) (\frac{\bs 1}{\bs s})'| \bs \beta^{(i)}, \bs \theta ] = \frac{1}{M}\sum_{j = 1}^M (\frac{\bs 1}{\bs s^{(j)}}) (\frac{\bs 1}{\bs s^{(j)}})'$; 
 	\item set $\bs \beta^{(i + 1)}= \text{argmin}_{\bs \beta} h(\bs y | \bs X,\bs \beta, \bs \phi ) + \frac{1}{2}\bs \beta' (\bs \Omega^{-1} \circ \widehat{\mathbb{E}}[(\frac{\bs 1}{\bs s}) (\frac{\bs 1}{\bs s})'| \bs \beta^{(i)}, \bs \theta ] ) \bs \beta$.
 \end{itemize}

Alternative posterior summaries can be obtained by simulating $M$ values $\bs \beta^{(1)},\dots, \bs \beta^{(M)}$ from the joint posterior distribution of $(\bs \beta, \bs s)$ given $\bs y$, $\bs X$, $\bs \phi$ and $\bs \theta$\color{black}, starting with \color{black} an initial value $\bs \beta^{(0)}$ and iteratively simulating $\bs s^{(j)}$ from the full conditional distribution of $\bs s$ given $\bs \beta^{(j - 1)}$ and $\bs \theta$ and simulating $\bs \beta^{(j)}$ from the full conditional distribution of $\bs \beta$ given $\bs y$, $\bs X$, $\bs s^{(j)}$ and $\bs \phi$. 
When the log-likelihood is quadratic or conditionally quadratic in $\bs \beta$, the full conditional distribution $p(\bs \beta | \bs s, \bs y, \bs X, \bs \phi)$ is a multivariate normal distribution.

\color{black}T\color{black}he full conditional distribution $p(\bs s | \bs \beta, \bs \Omega, \bs \theta)$ can be written as proportional to
\begin{align}\label{eq:fulls}
p(\bs s | \bs \beta, \bs \Omega, \bs \theta)\propto &\left(\prod_{j = 1}^p \left|s_{\color{black}j\color{black}}\right|^{-1}\right)\text{exp}\left\{-\frac{1}{2}\left(\bs 1/\bs s\right)'\left(\bs \Omega^{-1}\circ \left(\bs \beta \bs \beta'\right)\right)\left(\bs 1/\bs s\right)\right)p\left(\bs s | \bs \theta\right), 
\end{align}

\hspace{-6mm}where `$/$' is applied elementwise.
The choices $p(\bs s | \bs \theta)$ that yield the SPN, SNG and SPB models do not yield standard distributions when $\bs \Omega$ is not a diagonal matrix.

We simulate from the full conditional distribution \eqref{eq:fulls} using generalized elliptical slice sampling \citep{Nishihara2014}. 
First, \color{black}
we unify all three models by recognizing that simulating $\boldsymbol s$ according to \eqref{eq:fulls} is equivalent to simulating unconstrained $\boldsymbol s'$ according to
\begin{align}
p(\bs s' | \bs \beta, \bs \Omega, \bs \theta)\propto &\left(\prod_{j = 1}^p \left|s'_{\color{black}j\color{black}}\right|^{-1}\right)\text{exp}\left\{-\frac{1}{2}\left(\bs 1/f\left(\bs s'\right)\right)'\left(\bs \Omega^{-1}\circ \left(\bs \beta \bs \beta'\right)\right)\left(\bs 1/f\left(\bs s'\right)\right)\right)p\left(f\left(\bs s'\right) | \bs \theta\right) \nonumber
\end{align}
and setting $\boldsymbol s = f\left(\boldsymbol s'\right)$, where $f\left(\boldsymbol s'\right) = \boldsymbol s'$ when the SPN prior is used and $f\left(\boldsymbol s'\right) = \left|\boldsymbol s'\right|$ when the SNG or SPB priors are used. Such use of the absolute value transformation is common in the  generalized elliptical slice sampling literature \citep{Nishihara2014}.
Then 
\color{black}
we perform a change of variables from $\bs s\color{black}'\color{black}$ to $\bs s'\color{black}'\color{black} = \bs V^{-1/2} \left(\bs s\color{black}'\color{black} - \bs m\right)$\color{black}, where $\bs m$ and $\bs V$ may depend on the current values of $\bs \beta$, $\bs \Omega$, and $\bs \theta$ but not $\boldsymbol s$ and $\bs V^{-1/2}$ refers to the symmetric square root of the pseudoinverse $\bs V^{\dagger}$ of $\bs V$. 
 Applying a change of variables when simulating from conditional distributions is a standard practice, as seen in \citep{Murray2010}.
\color{black}
Simulating $\bs s$ according to \eqref{eq:fulls} is equivalent to simulating $\bs s'\color{black}'\color{black}$ according to 
\begin{align}\label{eq:fullsp}
p\left(\bs s'\color{black}'\color{black} | \bs \beta, \bs \Omega, \bs \theta\right) \propto&\left|\text{diag}\left\{\left| \bs V^{1/2}\bs s'\color{black}'\color{black} + \boldsymbol m\right| \right\}\right|^{-1} \times \\ \nonumber
&\text{exp}\left\{-\frac{1}{2}\left(\bs 1/\color{black}f\color{black}\left(\bs V^{1/2}\bs s'\color{black}'\color{black} + \boldsymbol m\right)\right)'\left(\bs \Omega^{-1}\circ \left(\bs \beta \bs \beta'\right)\right)\left(\bs 1/\color{black}f\color{black}\left(\bs V^{1/2}\bs s'\color{black}'\color{black} + \boldsymbol m\right)\right)\right) \times \\ \nonumber
&p\left(\color{black}f\left(\color{black}\bs V^{1/2}\bs s'\color{black}'\color{black} + \boldsymbol m\color{black}\right)\color{black} | \bs \theta\right)
\end{align}
\color{black}
and setting $\boldsymbol s = f\left(\boldsymbol V^{1/2} \boldsymbol s'' + \boldsymbol m \right)$,  
\color{black}
where `$/$' and $\left|\cdot \right|$ are applied elementwise.
In turn, simulating $\bs s'\color{black}'\color{black}$ according to \eqref{eq:fullsp} is equivalent to simulating four $p\times 1$ vectors $\bs r$, $\bs u$, and $\bs \pi$ according to
	\begin{align}\label{eq:ess}
	p\left(\bs s\color{black}''\color{black} = \bs u \text{sin}(\bs \pi) + \bs r \text{cos}(\bs \pi) | \bs \beta, \bs \Omega, \bs \theta\right)
	\prod_{j = 1}^p  \text{exp}\left\{-\frac{1}{2}\left(r_j^2 + u_j^2\right)\right\}/\text{exp}\left\{-\frac{s_j{'\color{black}'\color{black}}^2}{2}\right\}.
	\end{align}
	For fixed $\bs m$ \color{black} and \color{black} $\bs V$, iteration $i$ of a Gibbs sampler simulating from \eqref{eq:ess} is as follows:
	\begin{enumerate}
		\item Simulate $t^{(i)}_j \sim \text{normal}(0, 1)$ for $j = 1, \dots, p$.
		\item Set $\bs u^{(i)}= (\bs s\color{black}''\color{black}^{(i - 1)})\text{sin}(\bs \pi^{(i - 1)}) + \bs t^{(i)}\text{cos}(\bs \pi^{(i - 1)})$. 
		\item Set $\bs r^{(i)}= (\bs s\color{black}''\color{black}^{(i - 1)})\text{cos}(\bs \pi^{(i - 1)}) - \bs t^{(i)}\text{sin}(\bs \pi^{(i - 1)})$.
		\item Simulate $\pi^{(i)}_j$ according to \eqref{eq:ess} for $j = 1,\dots, p$ using univariate slice sampling.
		\item Set $\bs s\color{black}''\color{black}^{(i)} = \bs u^{(i)} \text{sin}(\bs \pi^{(i)}) + \bs r^{(i)} \text{cos}(\bs \pi^{(i)})$\color{black}.\color{black}
		\item Set $\bs s^{(i)} = \color{black}f\left(\color{black}\bs V^{1/2} \bs s'\color{black}'\color{black}^{(i)} + \bs m\color{black}\right)\color{black}$.
	\end{enumerate}
	\color{black}Univariate slice sampling is described in the appendix. \color{black}
	We can think of steps 1-4 as generating a proposal for $\bs s'\color{black}'\color{black}^{(i)}$ and step 5 as generating an adjustment for the proposal that reflects the relationship between the proposal distribution and the target full conditional distribution. \color{black} Validity of the proposed algorithm follows from \cite{Murray2010}. \color{black}
	The \color{black} vector \color{black} $\bs m$ and \color{black} matrix \color{black} $\bs V$ can be chosen to improve mixing and decrease computational burden. Choices of $\bs m$ and $\bs V$ that provide better approximations to the full conditional distribution $p(\bs s | \bs \beta, \bs \Omega, \bs \theta)$ are likely to result in better mixing, whereas choices of $\bs m$ and $\bs V$ that are easier to compute will improve the speed of the sampling process. Here, $\bs m$ and $\bs V$ are chosen to approximate the mode of $p(\bs s\color{black}'\color{black} | \bs \beta, \bs \Omega, \bs \theta)$ and \color{black} corresponding Hessian\color{black}. We compute $\bs m$, an approximate mode of  $p(\bs s\color{black}'\color{black} | \bs \beta, \bs \Omega, \bs \theta)$, by performing coordinate descent with a large convergence threshold and a small number of maximum iterations. We describe how $\bs m$ can be obtained via coordinate descent under the SHP priors in the appendix. 
	
	\subsubsection{Simulating from the Posterior Distribution under the SPN Prior}
	As noted in Section~\ref{sec:pri}, \color{black} simulation \color{black} from the joint posterior distribution under the SPN prior \color{black} is simple\color{black}.
	In the linear regression setting with known unit variance where $-h(\bs y | \bs X, \bs \beta )\propto_{\bs \beta} \frac{1}{2}(\bs \beta'\bs X'\bs X \bs \beta - 2\bs \beta'\bs X'\bs y)$, then 
\begin{align*}
\bs z |\bs X, \bs y, \bs s, \bs \Omega &\sim \text{normal}\left(\left(\left(\bs X'\bs X\right) \circ \left(\bs s \bs s'\right) + \bs \Omega^{-1}\right)^{-1}\left(\bs s \circ \left(\bs X'\bs y\right)\right),  \left(\left(\bs X'\bs X\right) \circ \left(\bs s \bs s'\right) + \bs \Omega^{-1}\right)^{-1}\right),\\
\bs s |\bs X, \bs y, \bs z, \bs \Psi &\sim \text{normal}\left(\left(\left(\bs X'\bs X\right) \circ \left(\bs z \bs z'\right) + \bs \Psi^{-1}\right)^{-1}\left(\bs z \circ \left(\bs X'\bs y\right)\right),  \left(\left(\bs X'\bs X\right) \circ \left(\bs z \bs z'\right) + \bs \Psi^{-1}\right)^{-1}\right).
\end{align*}

Both full conditional distributions are multivariate normal distributions. As a result, a straightforward Gibbs sampler can be used to simulate from the joint posterior distribution.

\subsection{Hyperparameter Estimation}\label{subsec:var}

In the previous subsection, we presented a general approach for simulating from the posterior distribution of $\bs \beta$ under the SPN, SNG and SPB priors given hyperparameters $\bs \Omega$ and, in the case of the SPN prior, $\bs \Psi$. 
\color{black}When the hyperparameters are unknown, one approach \color{black} is to \color{black}assume \color{black} prior distributions for $\bs \phi$, $\bs \Omega$, and/or $\bs \Psi$. For all three priors, a conjugate inverse-Wishart prior for $\bs \Omega$ is a natural choice. For the SPN prior, a conjugate inverse-Wishart prior for $\bs \Psi$ is likewise natural. 
\color{black} 
There are situations where conjugate inverse-Wishart priors may have undesirable properties \citep{Schuurman2016}, in which case alternative prior distributions for $\bs \Omega$ and, when the SPN prior is used, $\bs \Psi$, could be used. 
\color{black}

\color{black}Alternatively\color{black}, hyperparameter estimates can be obtained via maximum marginal likelihood estimation (MMLE) or the method of moments.
\color{black}W\color{black}e can \color{black}compute the MMLE \color{black} of the unknown variance components $\bs \Omega$ and, in the case of the SPN prior, $\bs \Psi$, using \color{black}a \color{black} Gibbs-within-EM algorithm as described in the appendix. 
However, maximum marginal likelihood estimation of hyperparameters can converge prohibitively slowly in practice \citep{Roy2016}. 
Furthermore, the Gibbs step can be prohibitively computationally demanding when the data are high dimensional.
Fortunately, method of moments type estimates of the unknown variance components can be obtained under the SNG and SPB priors for fixed $c$ and $q$ respectively, and under the symmetric sSPN prior, so long as $\bs y$ is linearly related to $\bs X \bs \beta$.
As noted in the Introduction, the prior moments are easy to compute under all three priors 
\begin{align*}
\mathbb{E}\left[\bs \beta \right] = \mathbb{E}\left[\bs s\right]\circ \mathbb{E}\left[\bs z\right]\text{ and } \bs \Sigma = \mathbb{E}\left[\bs s \bs s'\right] \circ \mathbb{E}\left[\bs z \bs z'\right].
\end{align*}
Furthermore, under the sSPN, SNG and SPB priors, the hyperparameters are correspond to second order moments of $\bs \beta$.
When $y$ is linearly related to $\bs X\bs \beta$, a positive semidefinite estimate of $\bs \Sigma$ can be obtained using methods from \cite{Perry2017}. 
Under the sSPN prior, estimates of $\bs \Omega$ and $\bs \Psi$ can be obtained from an estimate $\hat{\bs \Sigma}$ by projecting $\sqrt{|\hat{\bs \Sigma}|}$ and $\text{sign}\{\hat{\bs \Sigma}\}\sqrt{|\hat{\bs \Sigma}|}$ onto the positive semi-definite cone, where $\sqrt{\cdot}$, $|\cdot|$ and $\text{sign}\{\cdot \}$ are applied elementwise. Under the SNG and SPB priors, an estimate of $\bs \Omega$ can be obtained from an estimate $\hat{\bs \Sigma}$ by projecting $\hat{\bs \Sigma}/((1 - \mathbb{E}[s_j ]^2) \bs I_p + \mathbb{E}[s_j]^2\bs 1_p \bs 1_p')$  onto the positive semi-definite cone, where `$/$' is applied elementwise, $\mathbb{E}[s_j ]^2 = c^{-1}(\Gamma(c + 1/2)/\Gamma(c))^2$ under the SNG prior and $\mathbb{E}[s_j]^2 = (\pi/2)(\Gamma(2/q)/\sqrt{\Gamma(1/q)\Gamma(3/q)})^2$ under the SPB prior.

\subsection{\color{black}Prior Selection}\label{subsec:modsel}
\color{black}
Our focus is on understanding the properties of each novel prior and providing a method for using them in practice. 
However, we would be remiss to omit guidance on when SNG, SPB, and SPN priors should be used as opposed to more common existing alternatives, how to choose between the SNG, SPB, and SPN priors when their use is warranted, and, in the case of the SNG and SPB priors, how to choose $c$ or $q$. 

Ideally, our understanding of the scientific problem would inform our choice of prior. Alternatively, we recommend first choosing a criterion for selecting a prior that is appropriate to the specific application and goals of analysis, e.g. Deviance Information Criterion (DIC), the widely applicable information criterion (WAIC), log marginal likelihood, mean log conditional predictive ordinate (CPO)/leave-one-out cross validation error (LOO) \citep{Roos2011, Piironen2017,Vehtari2017}.  We then recommend performing an initial analysis that compares the chosen criterion under an independent normal, multivariate normal, and independent sparsity inducing prior of interest, e.g. a Laplace prior. If both the multivariate normal and independent sparsity inducing prior provide improved fit compared to an independent normal prior, we recommend considering the SPN, SNG and/or SPB priors for a range of values of $c$ and/or $q$ and selecting the prior that is best according to the chosen criterion. 
\color{black}

\section{Application}\label{sec:num}

In this section we return to the \color{black} subset of \color{black} BCI data discussed in Section~\ref{sec:intro}.
We assume the following model for the \color{black} indictors of whether or not the subject was viewing the target letter $\bs y$ and EEG measurements at time point $j$ and channel $k$, $\bs x_{jk}$ during the first 20 trials,\color{black}
\begin{align*}
y_i \stackrel{indep.}{\sim} \text{Binomial} ((1 + \text{exp}\{-\gamma_{jk} - \beta_{jk} x_{ijk} \})^{-1}),
\end{align*} 
where each $\gamma_{jk}$ is a time point and channel specific intercept and each $\beta_{jk}$ is a regression coefficient that describes the relationship between EEG measurements from time point $j$ and channel $k$ and whether or not the subject was viewing the target letter, for $j = 1, \dots, p_1$ and $k = 1, \dots, p_2$. 
Because simulating from the posterior distribution of $\bs \beta$ becomes increasingly computationally demanding as the number of unknown regression coefficients $\beta_{jk}$ increases and because we want to explore the results of using several different  SNG and SPB priors, we consider a reduced set of time points, specifically every eighth time point. This reduces $p_1$ from $208$ to $26$. We continue to consider all channels, $p_2 = 8$. \color{black} 
 
\color{black}Letting $\bs \beta = \text{vec}\left(\bs B\right)$ refer to a vectorized $p_1\times p_2$ matrix $\bs B$ with elements $b_{jk} = \beta_{jk}$, 
w\color{black}e assume that $\mathbb{E}\left[\bs \beta\right] = \bs 0$ and $\mathbb{V}\left[\bs \beta\right] = \bs \Sigma$, where $\bs \Sigma$ is separable, i.e.\ $\bs \Sigma = \bs \Sigma_2 \otimes \bs \Sigma_1$. The covariance matrices $\bs \Sigma_1$ and $\bs \Sigma_2$ characterize relationship of regression coefficients $\bs B$ over time and across channels, respectively. Because the scales of $\bs \Sigma_1$ and $\bs \Sigma_2$ are not separately identifiable, we assume that the diagonal elements of $\bs \Sigma_1$ are exactly equal to 1. 
We also assume that $\bs \Omega_1$ and and $\bs \Psi_1$ have autoregressive structures of order one, with $\omega_{1,ij} = \rho^{\left|i-j\right|}_{\bs \Omega}$ and $\psi_{1,ij} = \rho^{\left|i-j\right|}_{\bs \Psi}$. 
We note that the corresponding marginal variance $\bs \Sigma_1$ is autoregressive of order one under the SPN prior but not the SNG and SPB priors. If $\bs \Omega_1$ and $\bs \Psi_1$ have autoregressive structure of order one with parameters $\rho_{\bs \Omega}$ and $\rho_{\bs \Psi}$, then $\bs \Sigma_1$ has autoregressive structure of order one with parameter $\rho = \rho_{\bs \Omega} \rho_{\bs \Psi}$. In contrast, the matrix $\bs \Sigma_1 = \bs \Omega_1 \circ \mathbb{E}\left[\bs s\bs s'\right]$ does not have autoregressive structure of order one.

\color{black} We assume \color{black} $\rho_{\bs \Omega}\sim \text{beta}\left(\left(p_1 + 1\right)/2, \left(p_1 + 1\right)/2\right)$ and $\bs \Omega^{-1}_2 \sim \text{Wishart}(p_2 + 2, \kappa^{-1}\bs I_{p_2})$, where \color{black} $\kappa = \frac{1}{p_1p_2 - 1} \sum_{j = 1}^{p_1} \sum_{k = 1}^{p_2} \left(\hat{\beta}_{logit,jk} - \bar{\hat{\beta}}_{logit} \right)^2$, $\bar{\hat{\beta}}_{logit} =\frac{1}{p_1p_2} \sum_{j = 1}^{p_1} \sum_{k = 1}^{p_2} \hat{\beta}_{logit,jk}$, and $\hat{\beta}_{logit,jk}$ are the logistic regression estimates of $\beta_{jk}$ depicted in Figure~\ref{fig:descc}\color{black}. When using the SPN prior, we also assume $\bs \Psi^{-1}_2\sim \text{Wishart}(p_2 + 2, \bs I_{p_2})$ and 
$\rho_{\bs \Psi}\sim \text{beta}\left(\left(p_1 + 1\right)/2, \left(p_1 + 1\right)/2\right)$.  \color{black} We assume conjugate inverse-Wishart priors for convenience, as our primary goal is to compare estimates under the SNG, SPB, and SPN priors for $\bs \beta$. We use the latent variable representation of the logistic regression model introduced in \cite{Polson2013}, s\color{black}imulate from the full conditional distribution of $\bs \beta$ jointly, use the elliptical slice sampling procedure described in Section~\ref{sec:comp} to simulate from the full conditional distribution of $\bs s$, and use univariate slice sampling to simulate from the full conditional distributions of $\rho_{\bs \Omega}$ and $\rho_{\bs \Psi}$.
We simulate \color{black}20 chains of 13,500 \color{black} samples from the posterior distribution, discard the initial $1,000$ \color{black} from each chain \color{black} as burn-in, and thin by a factor of $5$. The remaining \color{black}50,000 \color{black} samples have minimum effective sample sizes \color{black} shown in the appendix. \color{black}

\begin{figure}[ht]
\centering
\includegraphics{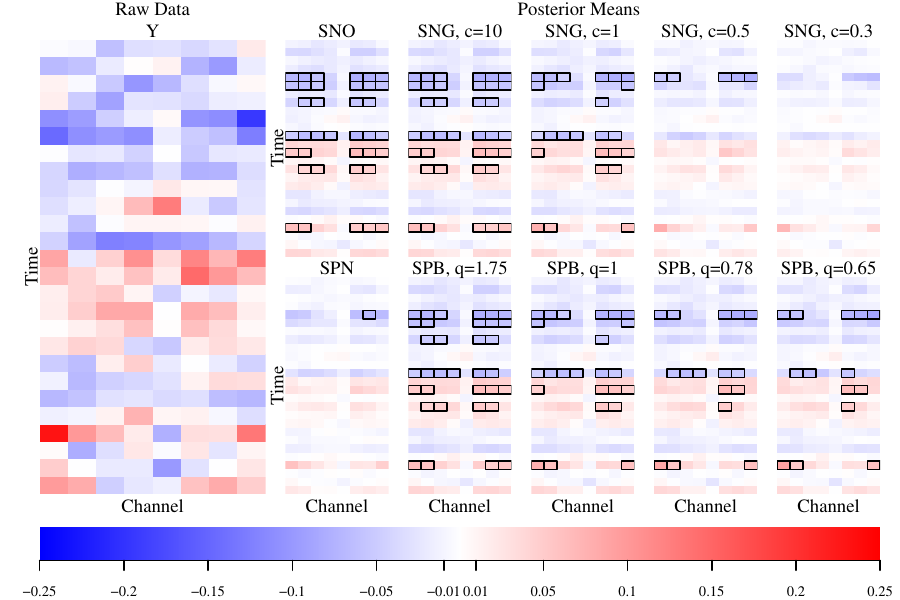}
\caption{Approximate posterior means of $\bs B$. \color{black} Boxes enclosing cells indicate approximate 90\% intervals that do not include zero.}
\label{fig:vbest}
\end{figure}

Figure~\ref{fig:vbest} shows posterior mean estimates of elements of $\bs B$ under a multivariate normal prior (SNO) and nine SHP priors. \color{black} Because shrinkage priors are often used to select nonzero elements of $\bs B$,  approximate 90\% intervals that do not include zero are also indicated. In cases where a sparse point estimate of $\bs B$ is desired, the methods of \cite{Hahn2015}, \cite{Li2017}, \cite{Piironen2020}, or \cite{Woody2021} can alternatively be used.  \color{black}
The estimated posterior means \color{black} and approximate 90\% intervals that do not include zero \color{black} display similar structure across channels and over time. \color{black}They also show \color{black} increasing shrinkage of individual elements of $\bs \beta$ as $c$ or $q \rightarrow 0$. Even the SNG \color{black} and SPB priors \color{black} with $c = 0.3$ and $q = 0.65$, which encourage sparsity very aggressively, produce estimates are structured, i.e. estimates of elements of $\bs B$ corresponding to the same channel or time point tend to be similar regardless of whether they are nearly equal to zero or large in magnitude. \color{black} In comparison to the normal prior, the more aggressive sparsity inducing priors facilitate identification of especially strong signals; for instance, they suggest that whether or not the subject is viewing the target letter is strongly correlated with EEG measurements at the fifth and twelfth time points on nearly all of the channels. \color{black}

\begin{figure}[ht]
\centering
\includegraphics{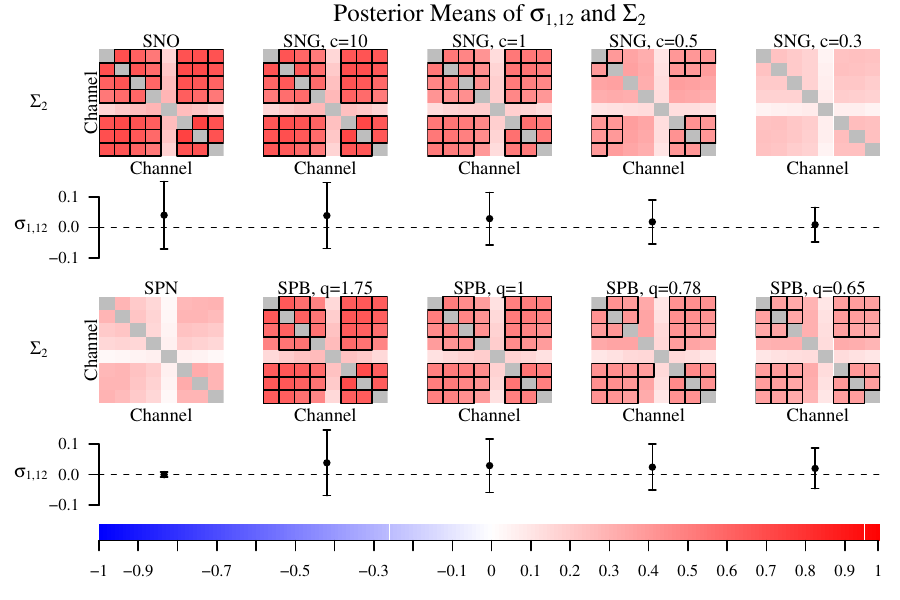}
\caption{Approximate posterior mean channel-by-channel correlation matrices $\bs \Sigma_2$\color{black}, with boxes enclosing cells with approximate 90\% intervals that do not include zero, \color{black} and correlations of consecutive time points $\sigma_{1,12}$ \color{black}with approximate 90\% intervals.}
\label{fig:vcorr}
\end{figure}
\color{black}

Figure~\ref{fig:vcorr} shows posterior mean estimates of channel-by-channel correlation matrices corresponding to $\bs \Sigma_2$ \color{black} and indications of approximate 90\% intervals that do not include zero as well as \color{black} posterior mean estimates of correlations across consecutive time points $\sigma_{1,i\left(i+1\right)}$ \color{black} and approximate 90\% intervals\color{black}.
\color{black}There is evidence of dependence, especially across channels. \color{black}
The amount of dependence across channels and over time under the SNG and SPB priors is decreasing as $c$ or $q\color{black} \rightarrow 0$\color{black}. \color{black} Surprisingly, the SPN prior estimates weak correlations despite being able to accommodate arbitrary $\bs \Sigma$. This may be due to especially strong shrinkage of elements of $\bs \Sigma$ under the assumed \color{black} prior distributions for the hyperparameters. \color{black}

\color{black}The available prior information, which suggests that $\bs B$ is sparse with strong signals observed across several channels at a subset of time points that roughly correspond to the expected timing of the P300 wave, can be used to assess the relative merits of the different priors.  The SNG and SPB priors with $c = 10$ and $q = 1.75$, respectively, produce estimates that are not sparse and so similar to the estimates produced under a normal prior that there is no added benefit relative to a multivariate normal prior. Meanwhile, the SNG prior with $c = 0.3$ and the SPN prior produce correlations that are not distinguishable from zero and thus provide no added benefit relative to an independent sparsity inducing prior and are implausibly sparse. This leaves the SNG and SPB priors with $c = 1$ and $q = 1$, which are equivalent and generalize the independent Laplace prior, the SNG prior with $c = 0.5$ and the SPB priors with $q = 0.78$, and $q=0.65$. The SNG prior with $c = 0.5$ misses signals that occur later in each trial that are likely to correspond to the known timing of the P300 wave. The SPB priors with $q = 0.78$ and $q = 0.65$ provide estimates that are qualitatively very similar to the estimates obtained under the SPB prior with $q = 1$.
Accordingly, we recommend the multivariate Laplace prior, which corresponds to the SNG prior with $c = 1$ and the SPB prior with $q = 1$; it is the simplest  SHP prior that produces estimates that align with the available scientific context. 

\begin{figure}[ht]
\centering
\includegraphics{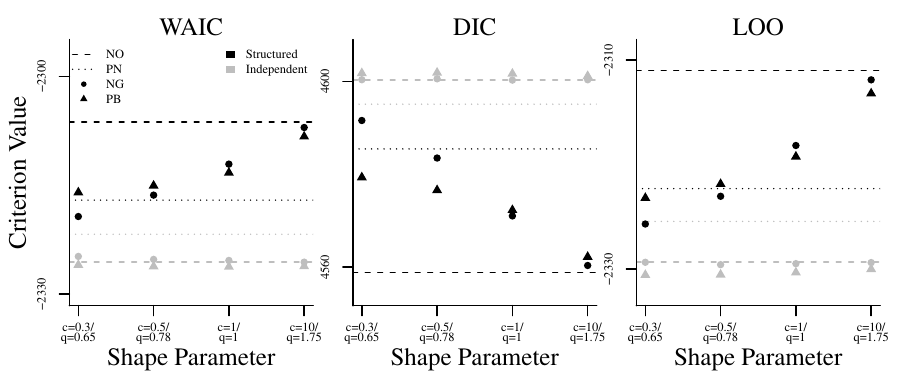}
\caption{\color{black} Approximate criteria for prior selection.}
\label{fig:mfit}
\end{figure}

For the purposes of demonstration, we also consider three criteria for prior selection; WAIC, DIC, and LOO. Approximations to these criteria are shown in Figure~\ref{fig:mfit}. Regardless of the criteria considered, the sparsity inducing independent product normal prior and the structured multivariate normal prior both outperform the the independent normal prior and therefore suggests that SHP priors are worth considering.
All structured priors outperform their independent counterparts. A detailed comparison of the estimates of $\bs B$ obtained under the structured priors we consider and their independent counterparts is included in the appendix.
Among structured priors, less sparsity inducing priors perform better and the multivariate normal prior performs best in this specific setting according to these criteria.

To conclude, we compare approximate 90\% intervals that do not include zero to ``ground truth'' 90\% confidence intervals that do not include zero based on simple logistic regression estimates computed from all 240 trials. Table~\ref{tab:compgt} shows that normal and nearly normal priors provide the best true positive rates, while the multivariate Laplace prior and SPB priors with $q < 1$ provide the best true positive rates relative to false positive rates.

\begin{table}
\centering 
\footnotesize
\color{black}
\begin{tabular}{c|cccccccccc}
& \multirow{2}{*}{SNO} & \multicolumn{4}{c}{SNG} & \multirow{2}{*}{SPN} & \multicolumn{4}{c}{SPB} \\
&  & $c = 10$ & $c = 1$ & $c = 0.5$ & $c = 0.3$ & & $q = 0.65$ & $q = 0.78$ & $q = 1$ & $q = 1.75$\\ \hline \hline
TP\% & 18.18 & 18.18 & 13.64 & 0.00 & 0.00 & 0.00 & 18.18 & 13.64 & 11.36 & 9.09 \\
FP\% & 17.68 & 17.68 & 10.98 & 3.05 & 0.00 & 0.61 & 15.85 & 10.98 & 6.71 & 6.10 \\ \hline 
\end{tabular}
\caption{\color{black}True positive (TP) and false positive (FP) rates of approximate 90\% intervals based on comparison to ``ground truth'' 90\% confidence intervals that do not include zero based on simple logistic regression estimates computed from all 240 trials.}
\label{tab:compgt}
\end{table}

\section{Discussion}\label{sec:conc}

We introduce the SHP class of novel structured shrinkage priors for regression coefficients \color{black} and show \color{black} that they can encourage both sparsity and structure, which can be difficult to simultaneously model using existing prior distributions. We provide a parsimonious and general approach to \color{black} simulation from the posterior distributions under SHP priors \color{black} based on elliptical slice sampling and demonstrate how SHP priors can improve interpretability of estimated regression coefficients relative to multivariate normal or independent priors.

\color{black}W\color{black}e have focused on the development of structured shrinkage prior distributions for regression coefficients, however the same distributions could be used to model errors as in \cite{Finegold2011} or as alternatives to Gaussian process\color{black}es\color{black}. We could also construct shrinkage priors for covariance matrices \color{black} by extending \cite{Daniels2002} or building on  the matrix-$F$ prior distribution for covariance matrices introduced in \cite{Mulder2018}, which assumes that $\bs \Sigma \stackrel{d}{=} \bs S \bs \Omega \bs S'$, where $\bs \Omega^{-1}$ has an inverse-Wishart distribution with scale $\bs I_p$ and degrees of freedom $\delta + k - 1$ and $\bs S\bs S'$ has a Wishart distribution with scale $\bs \Xi$ and degrees of freedom $\nu$.  We could
define distributions for  covariance matrices according to $\bs \Sigma \stackrel{d}{=} \text{diag}\left\{\bs s\right\} \bs \Omega \text{diag}\left\{\bs s\right\}$ for $\bs s \sim \text{normal}\left(\boldsymbol 0, \boldsymbol \Psi\right)$, $s^2_j \stackrel{i.i.d.}{\sim} \text{gamma}(c, c)$, or $s^{2}_j$  independently distributed according to a polynomially tilted positive $\alpha$-stable distribution with index of stability $\alpha = q/2$ and $\mathbb{E}[s^2_j] = 1$. \color{black} Also, the elliptical slice sampling procedure we use to construct a tractable Gibbs sampler for simulating from $p(\bs s | \bs \beta, \bs \theta)$ under the SPN, SNG and SPB priors can be to perform posterior inference under other novel structured generalizations of \color{black}other \color{black} shrinkage priors that can be represented using Hadamard products involving a normal random vector\color{black}, including the horseshoe and Dirichlet-Laplace priors \citep{Polson2010, Bhattacharya2015}.
Popularity of the horseshoe prior suggests that a structured generalization may be especially valuable.
Additional extensions might be to assess whether or not conditions for the convergence of elliptical slice sampling algorithms provided in \cite{Natarovskii2021} hold for the proposed elliptical slice sampling procedure and to improve scalability of the proposed elliptical slice sampling procedure with respect to $p$, the total number of penalized covariates. 
As shown in the appendix, seconds per posterior draw increases and minimum effective sample size decreases under these structured priors as more penalized covariates are used. The time costs of using structured priors as opposed to their independent counterparts increases with the number of penalized covariates and is greater for priors that encourage sparsity more aggressively.
\color{black}
\color{black} Another \color{black} extension to this work would treat $c$ and $q$ as unknown under the SNG and SPB priors, respectively\color{black}, and \color{black} perform maximum marginal likelihood estimation of $c$ or $q$ or assume prior distributions for $c$ and $q$.
\color{black}
Last, an extension might consider modifications of the proposed priors that continue to allow for structure in the presence of extreme elementwise sparsity, e.g. extensions of SNG and SPB priors  that allow for correlated scales and extensions of the SPN prior that can encourage shrinkage more aggressively by considering elementwise products of more than two normal vectors.
\color{black}

\section*{Acknowledgements}
This work was partially supported by NSF grants DGE-1256082 and DMS-1505136. 

\section*{Supplementary Materials}
Supplementary material available online includes proofs of all the propositions and additional numerical results.
A stand-alone package for implementing the methods described in this paper can be downloaded from \href{https://github.com/maryclare/sspcomp}{\texttt{https://github.com/maryclare/sspcomp}}.

\bibliography{MVBL.bib}

\begin{thebibliography}{}

\bibitem[\protect\citeauthoryear{Bhattacharya, Pati, Pillai, and
  Dunson}{Bhattacharya et~al.}{2015}]{Bhattacharya2015}
Bhattacharya, A., D.~Pati, N.~S. Pillai, and D.~B. Dunson (2015).
\newblock Dirichlet – laplace priors for optimal shrinkage.
\newblock {\em Journal of the American Statistical Association\/}~{\em 110},
  1479--1490.

\bibitem[\protect\citeauthoryear{Caron and Doucet}{Caron and
  Doucet}{2008}]{Caron2008}
Caron, F. and A.~Doucet (2008).
\newblock Sparse bayesian nonparametric regression.
\newblock {\em International Conference on Machine Learning\/}, 88--95.

\bibitem[\protect\citeauthoryear{Carvalho, Polson, and Scott}{Carvalho
  et~al.}{2010}]{Carvalho2010}
Carvalho, C.~M., N.~G. Polson, and J.~G. Scott (2010).
\newblock The horseshoe estimator for sparse signals.
\newblock {\em Biometrika\/}~{\em 97}, 465--480.

\bibitem[\protect\citeauthoryear{Daniels and Pourahmadi}{Daniels and
  Pourahmadi}{2002}]{Daniels2002}
Daniels, M.~J. and M.~Pourahmadi (2002).
\newblock Bayesian analysis of covariance matrices and dynamic models for
  longitudinal data.
\newblock {\em Biometrika\/}~{\em 89}, 553--566.

\bibitem[\protect\citeauthoryear{de~Brecht and Yamagishi}{de~Brecht and
  Yamagishi}{2012}]{DeBrecht2012}
de~Brecht, M. and N.~Yamagishi (2012).
\newblock Combining sparseness and smoothness improves classification accuracy
  and interpretability.
\newblock {\em NeuroImage\/}~{\em 60}, 1550--1561.

\bibitem[\protect\citeauthoryear{Dempster, Laird, and Rubin}{Dempster
  et~al.}{1977}]{Dempster1977}
Dempster, A.~P., N.~M. Laird, and D.~B. Rubin (1977).
\newblock Maximum likelihood from incomplete data via the em algorithm.
\newblock {\em Journal of the Royal Statistical Society. Series B\/}~{\em 39},
  1―38.

\bibitem[\protect\citeauthoryear{Devroye}{Devroye}{2009}]{Devroye2009}
Devroye, L. (2009).
\newblock Random variate generation for exponentially and polynomially tilted
  stable distributions.
\newblock {\em ACM Transactions on Modeling and Computer Simulation\/}~{\em
  19}, 1--20.

\bibitem[\protect\citeauthoryear{Finegold and Drton}{Finegold and
  Drton}{2011}]{Finegold2011}
Finegold, M. and M.~Drton (2011).
\newblock Robust graphical modeling of gene networks using classical and
  alternative t-distributions.
\newblock {\em Annals of Applied Statistics\/}~{\em 5}, 1057--1080.

\bibitem[\protect\citeauthoryear{Forney, Anderson, Davies, Gavin, Taylor, and
  Roll}{Forney et~al.}{2013}]{Forney2013}
Forney, E., C.~Anderson, P.~Davies, W.~Gavin, B.~Taylor, and M.~Roll (2013).
\newblock A comparison of eeg systems for use in p300 spellers by users with
  motor impairments in real-world environments.
\newblock {\em Proceedings of the Fifth International Brain-Computer Interface
  Meeting\/}.

\bibitem[\protect\citeauthoryear{Frank and Friedman}{Frank and
  Friedman}{1993}]{Frank1993}
Frank, I.~E. and J.~H. Friedman (1993).
\newblock A statistical view of some chemometrics regression tools.
\newblock {\em Technometrics\/}~{\em 35}, 109.

\bibitem[\protect\citeauthoryear{Griffin and Brown}{Griffin and
  Brown}{2010}]{Griffin2010}
Griffin, J.~E. and P.~J. Brown (2010).
\newblock Inference with normal-gamma prior distributions in regression
  problems.
\newblock {\em Bayesian Analysis\/}~{\em 5}, 171--188.

\bibitem[\protect\citeauthoryear{Griffin and Brown}{Griffin and
  Brown}{2012a}]{Griffin2012a}
Griffin, J.~E. and P.~J. Brown (2012a).
\newblock Competing sparsity: A hierarchical prior for sparse regression with
  grouped effects.
\newblock {\em Manuscript\/}.

\bibitem[\protect\citeauthoryear{Griffin and Brown}{Griffin and
  Brown}{2012b}]{Griffin2012}
Griffin, J.~E. and P.~J. Brown (2012b).
\newblock Structuring shrinkage: some correlated priors for regression.
\newblock {\em Biometrika\/}~{\em 99}, 481--487.

\bibitem[\protect\citeauthoryear{Hahn and Carvalho}{Hahn and
  Carvalho}{2015}]{Hahn2015}
Hahn, P.~R. and C.~M. Carvalho (2015, 1).
\newblock Decoupling shrinkage and selection in bayesian linear models: A
  posterior summary perspective.

\bibitem[\protect\citeauthoryear{Hoff}{Hoff}{2017}]{Hoff2016b}
Hoff, P.~D. (2017).
\newblock Lasso, fractional norm and structured sparse estimation using a
  hadamard product parametrization.
\newblock {\em Computational Statistics and Data Analysis\/}~{\em 115},
  186--198.

\bibitem[\protect\citeauthoryear{Kalli and Griffin}{Kalli and
  Griffin}{2014}]{Kalli2014}
Kalli, M. and Griffin (2014).
\newblock Time-varying sparsity in dynamic regression models.
\newblock {\em Journal of Econometrics\/}~{\em 178}, 779--793.

\bibitem[\protect\citeauthoryear{Kowal, Matteson, and Ruppert}{Kowal
  et~al.}{2017}]{Kowal2017}
Kowal, D.~R., D.~S. Matteson, and D.~Ruppert (2017).
\newblock Dynamic shrinkage processes.
\newblock {\em ArXiv preprint. arXiv:1707.00763\/}, 1--45.

\bibitem[\protect\citeauthoryear{Kyung, Gill, Ghosh, and Casella}{Kyung
  et~al.}{2010}]{Kyung2010}
Kyung, M., J.~Gill, M.~Ghosh, and G.~Casella (2010).
\newblock Penalized regression, standard errors, and bayesian lassos.
\newblock {\em Bayesian Analysis\/}~{\em 5}, 369--412.

\bibitem[\protect\citeauthoryear{Li and Pati}{Li and Pati}{2017}]{Li2017}
Li, H. and D.~Pati (2017, 3).
\newblock Variable selection using shrinkage priors.
\newblock {\em Computational Statistics and Data Analysis\/}~{\em 107},
  107--119.

\bibitem[\protect\citeauthoryear{Makeig, Kothe, Mullen, Bigdely-Shamlo, Zhang,
  and Kreutz-Delgado}{Makeig et~al.}{2012}]{Makeig2012}
Makeig, S., C.~Kothe, T.~Mullen, N.~Bigdely-Shamlo, Z.~Zhang, and
  K.~Kreutz-Delgado (2012).
\newblock Evolving signal processing for brain – computer interfaces.
\newblock {\em Proceedings of the IEEE\/}~{\em 100}, 1567--1584.

\bibitem[\protect\citeauthoryear{Mulder and Pericchi}{Mulder and
  Pericchi}{2018}]{Mulder2018}
Mulder, J. and L.~R. Pericchi (2018).
\newblock The matrix-f prior for estimating and testing covariance matrices.
\newblock {\em Bayesian Analysis\/}~{\em 13}, 1189--1210.

\bibitem[\protect\citeauthoryear{Murray, Adams, and Mackay}{Murray
  et~al.}{2010}]{Murray2010}
Murray, I., R.~P. Adams, and D.~J.~C. Mackay (2010).
\newblock Elliptical slice sampling.
\newblock {\em Journal of Machine Learning Research: WC\&P\/}~{\em 9},
  541--548.

\bibitem[\protect\citeauthoryear{Natarovskii, Rudolf, and Sprungk}{Natarovskii
  et~al.}{2021}]{Natarovskii2021}
Natarovskii, V., D.~Rudolf, and B.~Sprungk (2021).
\newblock Geometric convergence of elliptical slice sampling.
\newblock {\em Proceedings of the 38th International Conference on Machine
  Learning\/}~{\em 139}, 7969--7978.

\bibitem[\protect\citeauthoryear{Neal}{Neal}{2003}]{Neal2003}
Neal, R.~M. (2003).
\newblock Slice sampling.
\newblock {\em Annals of Statistics\/}~{\em 31}, 758--767.

\bibitem[\protect\citeauthoryear{Ng and Abugharbieh}{Ng and
  Abugharbieh}{2011}]{Ng2011}
Ng, B. and R.~Abugharbieh (2011).
\newblock Modeling spatiotemporal structure in fmri brain decoding using
  generalized sparse classifiers.
\newblock {\em IEEE International Workshop on Pattern Recognition in
  NeuroImaging\/}, 65--68.

\bibitem[\protect\citeauthoryear{Nishihara, Murray, and Adams}{Nishihara
  et~al.}{2014}]{Nishihara2014}
Nishihara, R., I.~Murray, and R.~P. Adams (2014).
\newblock Parallel mcmc with generalized elliptical slice sampling.
\newblock {\em Journal of Machine Learning Research\/}~{\em 15}, 2087--2112.

\bibitem[\protect\citeauthoryear{Park and Casella}{Park and
  Casella}{2008}]{Park2008}
Park, T. and G.~Casella (2008).
\newblock The bayesian lasso.
\newblock {\em Journal of the American Statistical Association\/}~{\em 103},
  681--686.

\bibitem[\protect\citeauthoryear{Perry}{Perry}{2017}]{Perry2017}
Perry, P.~O. (2017).
\newblock Fast moment-based estimation for hierarchical models.
\newblock {\em Journal of the Royal Statistical Society. Series B\/}~{\em 79},
  267--291.

\bibitem[\protect\citeauthoryear{Piironen, Paasiniemi, and Vehtari}{Piironen
  et~al.}{2020}]{Piironen2020}
Piironen, J., M.~Paasiniemi, and A.~Vehtari (2020).
\newblock Projective inference in high-dimensional problems: Prediction and
  feature selection.
\newblock {\em Electronic Journal of Statistics\/}~{\em 14}, 2155--2197.

\bibitem[\protect\citeauthoryear{Piironen and Vehtari}{Piironen and
  Vehtari}{2017}]{Piironen2017}
Piironen, J. and A.~Vehtari (2017).
\newblock Comparison of bayesian predictive methods for model selection.
\newblock {\em Statistics and Computing\/}~{\em 27}, 711--735.

\bibitem[\protect\citeauthoryear{Polson and Scott}{Polson and
  Scott}{2010}]{Polson2010}
Polson, N.~G. and J.~G. Scott (2010).
\newblock Shrink globally, act locally: Bayesian sparsity and regularization.

\bibitem[\protect\citeauthoryear{Polson, Scott, and Windle}{Polson
  et~al.}{2013}]{Polson2013}
Polson, N.~G., J.~G. Scott, and J.~Windle (2013).
\newblock Bayesian inference for logistic models using pólya-gamma latent
  variables.
\newblock {\em Journal of the American Statistical Association\/}~{\em 108},
  1339--1349.

\bibitem[\protect\citeauthoryear{Polson, Scott, and Windle}{Polson
  et~al.}{2014}]{Polson2014}
Polson, N.~G., J.~G. Scott, and J.~Windle (2014).
\newblock The bayesian bridge.
\newblock {\em Journal of the Royal Statistical Society. Series B: Statistical
  Methodology\/}~{\em 76}, 713--733.

\bibitem[\protect\citeauthoryear{Roos and Held}{Roos and Held}{2011}]{Roos2011}
Roos, M. and L.~Held (2011).
\newblock Sensitivity analysis in bayesian generalized linear mixed models for
  binary data.
\newblock {\em Bayesian Analysis\/}~{\em 6}, 259--278.

\bibitem[\protect\citeauthoryear{Roy, Reich, Guinness, Shinohara, and
  Staicu}{Roy et~al.}{2021}]{Roy2021}
Roy, A., B.~J. Reich, J.~Guinness, R.~T. Shinohara, and A.-M. Staicu (2021).
\newblock Spatial shrinkage via the product independent gaussian process prior.
\newblock {\em Journal of Computational and Graphical Statistics\/}~{\em 0},
  1--13.

\bibitem[\protect\citeauthoryear{Roy and Chakraborty}{Roy and
  Chakraborty}{2016}]{Roy2016}
Roy, V. and S.~Chakraborty (2016).
\newblock Selection of tuning parameters, solution paths and standard errors
  for bayesian lassos.
\newblock {\em Bayesian Analysis\/}~{\em 12}, 753--778.

\bibitem[\protect\citeauthoryear{Schuurman, Grasman, and Hamaker}{Schuurman
  et~al.}{2016}]{Schuurman2016}
Schuurman, N.~K., R.~P. Grasman, and E.~L. Hamaker (2016, 5).
\newblock A comparison of inverse-wishart prior specifications for covariance
  matrices in multilevel autoregressive models.
\newblock {\em Multivariate Behavioral Research\/}~{\em 51}, 185--206.

\bibitem[\protect\citeauthoryear{Sharma}{Sharma}{2013}]{Sharma2013}
Sharma, N. (2013).
\newblock Single-trial p300 classification with lda and neural networks.

\bibitem[\protect\citeauthoryear{Simon, Friedman, Hastie, and Tibshirani}{Simon
  et~al.}{2013}]{Simon2013}
Simon, N., J.~Friedman, T.~Hastie, and R.~Tibshirani (2013).
\newblock A sparse-group lasso.
\newblock {\em Journal of Computational and Graphical Statistics\/}~{\em 22},
  231--245.

\bibitem[\protect\citeauthoryear{Styan}{Styan}{1973}]{Styan1973}
Styan, P.~H. (1973).
\newblock Hadamard products and multivariate statistical analysis.
\newblock {\em Linear Algebra and Its Applications\/}~{\em 6}, 217--240.

\bibitem[\protect\citeauthoryear{Tibshirani}{Tibshirani}{1996}]{Tibshirani1996}
Tibshirani, R. (1996).
\newblock Regression shrinkage and selection via the lasso.
\newblock {\em Journal of the Royal Statistical Society: Series B (Statistical
  Methodology)\/}~{\em 58}, 267--288.

\bibitem[\protect\citeauthoryear{van Gerven, Cseke, Oostenveld, and Heskes}{van
  Gerven et~al.}{2009}]{vanGerven2009}
van Gerven, M., B.~Cseke, R.~Oostenveld, and T.~Heskes (2009).
\newblock Bayesian source localization with the multivariate laplace prior.
\newblock {\em Advances in Neural Information Processing Systems\/}~{\em 22},
  1--9.
\newblock Example of using<br/><br/>It's sort of related to the normal product
  prior<br/><br/>Perform approximate inference.

\bibitem[\protect\citeauthoryear{van Gerven, Cseke, de~Lange, and Heskes}{van
  Gerven et~al.}{2010}]{vanGerven2010}
van Gerven, M. A.~J., B.~Cseke, F.~P. de~Lange, and T.~Heskes (2010).
\newblock Efficient bayesian multivariate fmri analysis using a sparsifying
  spatio-temporal prior.
\newblock {\em Neuroimage\/}~{\em 50}, 150--161.
\newblock Very relevant!<br/><br/>Approximate inference.

\bibitem[\protect\citeauthoryear{Vehtari, Gelman, and Gabry}{Vehtari
  et~al.}{2017}]{Vehtari2017}
Vehtari, A., A.~Gelman, and J.~Gabry (2017).
\newblock Practical bayesian model evaluation using leave-one-out
  cross-validation and waic.
\newblock {\em Statistics and Computing\/}~{\em 27}, 1413--1432.

\bibitem[\protect\citeauthoryear{Wolpaw and Wolpaw}{Wolpaw and
  Wolpaw}{2012}]{Wolpaw2012}
Wolpaw, J. and E.~W. Wolpaw (2012).
\newblock {\em Brain-Computer Interfaces: Principles and Practice}.
\newblock Oxford University Press.

\bibitem[\protect\citeauthoryear{Woody, Carvalho, and Murray}{Woody
  et~al.}{2021}]{Woody2021}
Woody, S., C.~M. Carvalho, and J.~S. Murray (2021).
\newblock Model interpretation through lower-dimensional posterior
  summarization.
\newblock {\em Journal of Computational and Graphical Statistics\/}~{\em 30},
  144--161.

\bibitem[\protect\citeauthoryear{Wu, Park, Koyejo, and Pillow}{Wu
  et~al.}{2014}]{Wu2014}
Wu, A., M.~Park, O.~Koyejo, and J.~W. Pillow (2014).
\newblock Sparse bayesian structure learning with dependent relevance
  determination prior.
\newblock {\em Advances in Neural Information Processing Systems\/},
  1628--1636.

\bibitem[\protect\citeauthoryear{Yuan and Lin}{Yuan and Lin}{2006}]{Yuan2006}
Yuan, M. and Y.~Lin (2006).
\newblock Model selection and estimation in regression with grouped variables.
\newblock {\em Journal of the Royal Statistical Society: Series B (Statistical
  Methodology)\/}~{\em 68}, 49--67.

\bibitem[\protect\citeauthoryear{Zhao, Gao, Mukherjee, and Engelhardt}{Zhao
  et~al.}{2016}]{Zhao2016}
Zhao, S., C.~Gao, S.~Mukherjee, and B.~E. Engelhardt (2016).
\newblock Bayesian group factor analysis with structured sparsity.
\newblock {\em Journal of Machine Learning Research\/}~{\em 17}, 1--47.

\end{thebibliography}
\bibliographystyle{chicago}

\clearpage

\title{Supplementary Material for \\
``Structured Shrinkage Priors''}
\author{}
\date{}
\maketitle

\begin{appendix}

\section{\color{black}Exploratory Data\color{black}}

\begin{figure}[ht]
\begin{center}
\hspace{-20mm}
\begin{subfigure}{.8\textwidth}
\centering
\includegraphics[width=5in]{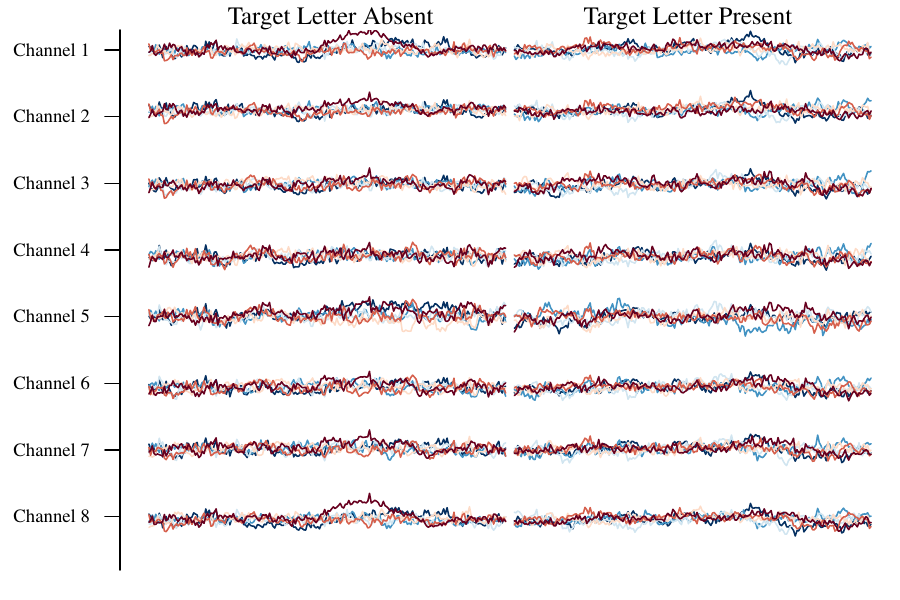}
\vspace{-10mm} \\
\caption{}
\label{fig:desc:a}
\end{subfigure} \hspace{-10mm}
\begin{subfigure}{.15\textwidth}
\centering \vspace{22.5mm} 
\includegraphics[width=1.4in]{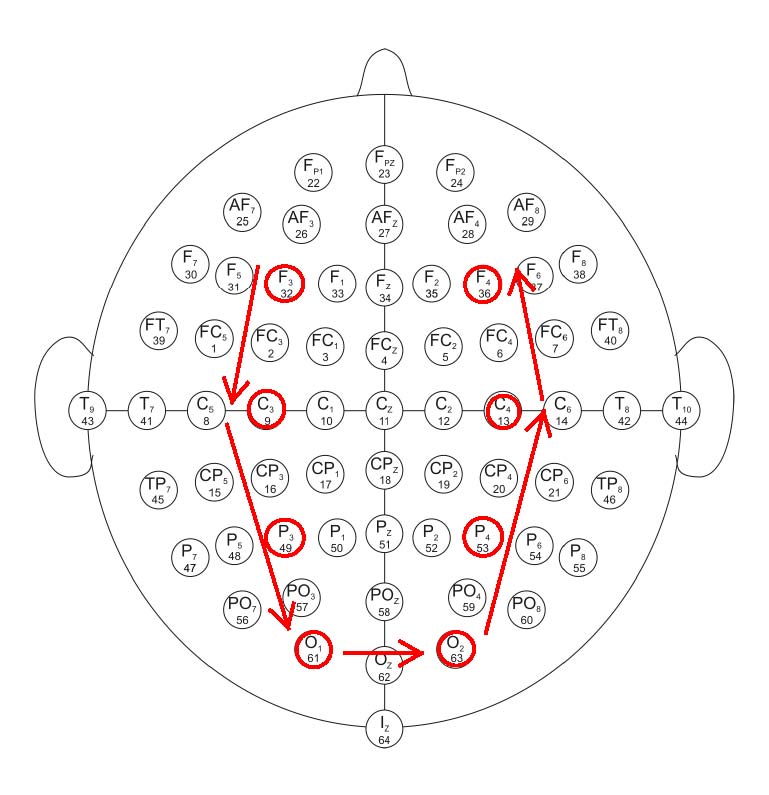}\\ \vspace{15mm}
\caption{}
\label{fig:desc:b}
\end{subfigure} 
\end{center}\vspace{-10mm}
\caption{Panel (a) shows a subset of single-subject P300 speller data. Lines represents trials, i.e.\ rows $\bs x_i$. Trials are plotted separately by whether or not the target letter was being shown during the trial. Panel (b) shows the locations of EEG sensors on the skull reprinted and modified from \cite{Sharma2013}, with sensors included in our analysis highlighted in red. Arrows indicate the order of the channels as they appear in the data.}
\label{fig:desc}
\end{figure}

\section{Relationship Between $\bs \Omega$, $\bs \Psi$ and Fourth-Order Prior Cross Moments}

Let $\bs C_{\bs \Omega}$ be the correlation matrix corresponding to $\bs \Omega$, and recall that parametrize $\bs \Psi$ as having diagonal elements equal to $1$.
Off-diagonal elements of $\bs C_{\bs \Omega}$ and $\bs \Psi$ determine fourth-order cross moments of elements of $\bs \beta$. Recalling that the marginal covariance matrix $\bs \Sigma = \bs \Omega \circ \bs \Psi$ and letting $\bs C_{\Sigma}$ be the correlation matrix corresponding to $\bs \Sigma$, we have
\begin{align*}
\frac{\mathbb{E}\left[\beta^2_j \beta_k^2 \right]}{\sigma_{jj}\sigma_{kk}} &= 1 + 2c_{\Omega,jk}^2 + 2c_{\Psi,jk}^2  + 4c_{\Sigma, jk}^2&& j \neq k  \\
\frac{\mathbb{E}\left[\beta^2_j \beta_k \beta_l \right]}{\sigma_{jj}\sqrt{\sigma_{kk}\sigma_{ll}}} &= c_{\Sigma,kl} + 4c_{\Sigma,jk}c_{\Sigma,jl} + 2c_{\Omega,kl}c_{\Psi,jk}c_{\Psi,jl} + 2c_{\Psi,kl} c_{\Omega,jk}c_{\Omega,jl}  && j \neq k\text{, }j \neq l\text{, } k\neq l. 
\end{align*}
We can see that these fourth-order cross moments depend on the values of \emph{both} $c_{\Omega,jk}$ and $c_{\Psi,jk}$, in addition to their product $c_{\Sigma,jk}$. 

\section{Generalized Gamma Rate Mixture Representation of Polynomially Tilted Positive $\alpha$-Stable Random Variables}
Following \cite{Devroye2009}, we can write polynomially titled positive $\alpha$-stable $s^2_j$ as: 
\begin{align*}
&s^2_j \stackrel{d}{=} \frac{\Gamma\left(\frac{1}{2\alpha}\right)\xi_j^{\frac{1 - \alpha}{\alpha}}}{2\Gamma\left(\frac{3}{2\alpha}\right)}\text{, } \xi_j \stackrel{i.i.d.}{\sim} \text{gamma}\left(\text{shape}=\frac{1 + \alpha}{2\alpha}, \text{rate}=f\left(\delta_j | \alpha\right)\right)\text{ and } \\
&p\left(\delta_j | \alpha\right)\propto f\left(\delta_j | \alpha \right)^{\frac{\alpha - 1}{2\alpha}},
\end{align*}
where $f(\delta_j | \alpha ) = \text{sin}(\alpha\delta_j)^{\frac{\alpha}{1 - \alpha}}\text{sin}((1 - \alpha)\delta_j)/\text{sin}(\delta_j)^{\frac{1}{1 - \alpha}}$ and $\delta_j \in (0, \pi)$. 
The density $f(\delta_j | \alpha )$ is non-standard, however \cite{Devroye2009} provides a following method for simulating from $\delta_j$ from this distribution.

\section{Univariate Marginal Distributions} 

Intuitively, it is clear from the stochastic representation that the marginal distributions are the same as the corresponding univariate shrinkage prior. We can show this directly as follows. The joint marginal prior distribution of $\bs \beta$ is 
	\begin{align*}
	p\left(\bs \beta\right) &= \int p\left(\bs s\right)p\left(\bs \beta/\bs s\right) \left(\prod_{j = 1}^p \frac{1}{\left|s_j\right|}\right) d s_1 \dots d s_p
	\end{align*}
	Then $p(\beta_1)$ is given by
	\begin{align*}
	p\left(\beta_1\right) &= \int p\left(\beta_1/s_1\right)p\left(\bs \beta_{-1}/\bs s_{-1} | \beta_1/s_1\right)\left(\prod_{j = 1}^p \frac{p\left(s_j\right)}{\left|s_j\right|}\right) ds_1 \dots d s_p d \beta_2\dots d\beta_p \\
	&= \int p\left(\beta_1/s_1\right)p\left(s_1\right)/\left|s_1\right| \underbrace{\left(\int p\left(\bs \beta_{-1}/\bs s_{-1} | \beta_1/s_1\right)\left(\prod_{j = 2}^p \frac{p\left(s_j\right)}{\left|s_j\right|}\right) ds_2 \dots d s_pd\beta_2\dots d\beta_p\right)}_{(*)} d s_1.
	\end{align*}
	The term $(*)$ is equal to $\int p(\bs \beta_{-1} | \beta_1/s_1)d\beta_2\dots d\beta_p$. This is the integral of a density, and accordingly $(*) = 1$ and $p(\beta_1) = \int p(\beta_1/s_1)p(s_1)/|s_1| d s_1$.

\section{Proofs of Propositions} 

\subsection{Propositions 2.1 and 2.3}
First, we prove the following lemma:
\begin{lemma}\label{lemma:infSPNp}
For $\alpha > 0$ and $\gamma \in \mathbb{R}$, $\int_{-\infty}^\infty \frac{1}{|s|} \text{exp}\{-\alpha (s^2 - \gamma s)\}ds = +\infty$.
\end{lemma}

First let's consider this integral when $\gamma > 0$:
\begin{align*}
\int_{-\infty}^{\infty} \frac{1}{\left|s\right|}\text{exp}\left\{-\alpha \left(s^2 - \gamma s\right) \right\} ds  = &\int_{-\infty}^{0} -\frac{1}{s}\text{exp}\left\{-\alpha s \right\}^{s - \gamma} ds + \\
&\int_{0}^{\gamma} \frac{1}{s}\text{exp}\left\{-\alpha s  \right\}^{s - \gamma} ds +  \int_{\gamma}^{\infty} \frac{1}{s}\text{exp}\left\{-\alpha s \right\}^{s - \gamma} ds.
\end{align*}
Because the integrand is nonnegative for all $s$, if \emph{any} of these terms evaluate to $+\infty$, the entire integral evaluates to $+\infty$. Let's examine the middle integral. Note that over this range, $s - \gamma \leq 0$, $\text{exp}\{-\alpha s \} \leq 1$ and accordingly, $\text{exp}\{-\alpha s\}^{s - \gamma} \geq 1$.
\begin{align*}
\int_{0}^{\gamma} \frac{1}{s}\text{exp}\left\{-\alpha s \right\}^{s - \gamma} ds &\geq  \int_{0}^{\gamma} \frac{1}{s} ds \\
&= \left(\text{ln}\left(\gamma\right) - \text{lim}_{a \rightarrow 0+}\text{ln}\left(a\right)\right)  = +\infty.
\end{align*}

Now let's consider the same integral when $\gamma < 0$:
\begin{align*}
\int_{-\infty}^{\infty} \frac{1}{\left|s\right|}\text{exp}\left\{-\alpha \left(s^2 - \gamma s\right) \right\} ds  = &\int_{-\infty}^{\gamma} -\frac{1}{s}\text{exp}\left\{-\alpha s \right\}^{s - \gamma} ds + \\
&\int_{\gamma}^{0} -\frac{1}{s}\text{exp}\left\{-\alpha s  \right\}^{s - \gamma} ds +  \int_{0}^{\infty} \frac{1}{s}\text{exp}\left\{-\alpha s \right\}^{s - \gamma} ds.
\end{align*}
Again, if \emph{any} of these terms evaluate to $+\infty$, the entire integral evaluates to $+\infty$. Again, let's consider the middle term. Over this interval, $\text{exp}\{-\alpha s \} \geq 1$ and $s - \gamma \geq 0$. It follows that $\text{exp}\{-\alpha s \}^{s - \gamma} \geq 1$ and:
\begin{align*}
\int_{\gamma}^{0} -\frac{1}{s}\text{exp}\left\{-\alpha s  \right\}^{s - \gamma} ds &\geq \int_{\gamma}^{0} -\frac{1}{s} ds \\
&= \int_{0}^{-\gamma} \frac{1}{s} ds = +\infty.
\end{align*}

Now we'll consider one last case where $\gamma = 0$:
\begin{align*}
\int_{-\infty}^{\infty} \frac{1}{\left|s\right|}\text{exp}\left\{-\alpha s^2 \right\} ds  = &\int_{-\infty}^{0} -\frac{1}{s}\text{exp}\left\{-\alpha s^2 \right\} ds +  \int_{0}^{1/\sqrt{\alpha}} \frac{1}{s}\text{exp}\left\{-\alpha s^2 \right\} ds + \int_{1/\sqrt{\alpha}}^{\infty} \frac{1}{s}\text{exp}\left\{-\alpha s^2 \right\} ds.
\end{align*}
As in the previous cases, \emph{any} of these terms evaluate to $+\infty$, the entire integral evaluates to $+\infty$.
Examining the middle term one last time, we have:
\begin{align*}
\int_{0}^{1/\sqrt{\alpha}} \frac{1}{s}\text{exp}\left\{-\alpha s^2 \right\} ds &\geq \int_{0}^{1/\sqrt{\alpha}} \frac{1}{s}\text{exp}\left\{-\sqrt{\alpha} s \right\} ds \\
&\geq \int_{0}^{1/\sqrt{\alpha}} \frac{1 -\sqrt{\alpha} s}{s} ds \\
&= \int_{0}^{1/\sqrt{\alpha}} \frac{1}{s} ds - \int_{0}^{1/\sqrt{\alpha}} \alpha ds \\
&= \text{ln}\left(1/\sqrt{\alpha}\right) - \text{lim}_{a\rightarrow 0+}\text{ln}\left(a\right)  -  \sqrt{\alpha} = +\infty.
\end{align*}

\subsubsection{Proof of Proposition 2.1:}
Now we can evaluate the marginal density $p(\bs \beta = \bs b | \bs \Omega, \bs \Psi)$ when $b_j = 0$ for some $j \in \{1, \dots, p\}$. Without loss of generality, set $\beta_1 = 0$. Letting  $\bs b_{-1}$ and $\bs s_{-1}$ refer to the vectors $\bs b$ and $\bs s$ each with the first element removed, $(\bs \Omega^{-1})_{-1,-1}$ be the matrix $\bs \Omega^{-1}$ with the first row and column removed, $(\bs \Psi^{-1})_{-1,-1}$ be the matrix $\bs \Psi^{-1}$ with the first row and column removed and $(\bs \Psi^{-1})_{-1,1}$ be the first column of $\bs \Psi^{-1}$ excluding the first element. 
For $\bs b = (0,\bs b_{-1})$,
\footnotesize
\begin{align*}
p\left(\left(0, \bs b_{-1}\right) | \bs \Psi, \bs \Omega \right) &\propto \int \frac{1}{\prod_{i = 1}^p\left|s_i\right|}\text{exp}\left\{-\frac{1}{2} \left(\bs b'\text{diag}\left\{\bs 1/\bs s\right\}\bs \Omega^{-1} \text{diag}\left\{\bs 1/\bs s\right\}\bs b + \bs s'\bs \Psi^{-1}\bs s\right) \right\} d\bs s \\
= \int &\frac{1}{\prod_{2 = 1}^p\left|s_i\right|}\text{exp}\left\{-\frac{1}{2} \left(\bs b_{-1}'\text{diag}\left\{\bs 1/\bs s_{-1}\right\}\left(\bs \Omega^{-1}\right)_{-1,-1} \text{diag}\left\{\bs 1/\bs s_{-1}\right\}\bs b_{-1} + \bs s_{-1}'\left(\bs \Psi^{-1}\right)_{-1,-1}\bs s_{-1}\right) \right\} \\
&\underbrace{\int_{-\infty}^{\infty} \frac{1}{\left|s_1\right|}\text{exp}\left\{-\frac{1}{2}\left(s^2_1 \left(\bs \Psi^{-1}\right)_{11}  - 2s_1 \left(\bs \Psi^{-1} \right)_{-1,1}'\bs s_{-1}\right) \right\}ds_1}_{(*)} d\bs s_{-1}.
\end{align*}
\normalsize
Applying Lemma~\ref{lemma:infSPNp}, the term denoted by $(*)$ evaluates to $+\infty$ for every value of $\bs s_{-1}$.

\subsubsection{Proof of Proposition 2.3:} Proposition 2.3 follows from Proposition 2.1 and Lemma~\ref{lemma:infSPNp}. Again, letting $\beta_1 = 0$, for $\bs \beta = (0, \bs \beta_{-1})$, 
\footnotesize
\begin{align*}
p\left( \beta_1 = 0 | \bs \Psi, \bs \Omega \right) &\propto \int \frac{1}{\prod_{i = 1}^p\left|s_i\right|}\text{exp}\left\{-\frac{1}{2} \left(\bs \beta'\text{diag}\left\{\bs 1/\bs s\right\}\bs \Omega^{-1}\text{diag}\left\{\bs 1/\bs s \right\}\bs \beta + \bs s'\bs \Psi^{-1}\bs s\right) \right\} d\bs s d\bs \beta_{-1} \\
= \int &\frac{1}{\prod_{2 = 1}^p\left|s_i\right|}\text{exp}\left\{-\frac{1}{2} \left(\bs \beta_{-1}'\text{diag}\left\{\bs 1/\bs s_{-1}\right\}\left(\bs \Omega^{-1}\right)_{-1,-1} \text{diag}\left\{\bs 1/\bs s_{-1}\right\}\bs \beta_{-1} + \bs s_{-1}'\left(\bs \Psi^{-1}\right)_{-1,-1}\bs s_{-1}\right) \right\} \\
&\underbrace{\int_{-\infty}^{\infty} \frac{1}{\left|s_1\right|}\text{exp}\left\{-\frac{1}{2}\left(s^2_1 \left(\bs \Psi^{-1}\right)_{11}  - 2s_1 \left(\bs \Psi^{-1} \right)_{-1,1}'\bs s_{-1}\right) \right\}ds_1}_{(*)} d\bs s_{-1}d\bs \beta_{-1}.
\end{align*}
\normalsize

Again, $(*)$ evaluates to $+\infty$ for any value of $\bs s_{-1}$ and does not depend at all on $\bs \beta_{-1}$.

\subsection{Proofs of Propositions 2.2 and 2.4} 

First, we prove the following lemma:
\begin{lemma}\label{lemma:infstng}
For $0 < c < 1/2$, $\int_{0}^\infty (s^2)^{c -1/2 - 1} \text{exp}\{-cs^2 \} ds^2 = +\infty$.
\end{lemma}

We can break the integral into two nonnegative components, as follows:
\begin{align*}
\int_{0}^\infty (s^2)^{c -1/2 - 1} \text{exp}\left\{-cs^2 \right\} ds^2 = \int_{0}^{1/c} (s^2)^{c -1/2 - 1} \text{exp}\left\{-cs^2 \right\} ds^2 + \int_{1/c}^{\infty} (s^2)^{c -1/2 - 1} \text{exp}\left\{-cs^2 \right\} ds^2.
\end{align*}

Now let's examine the first component. When $s^2 < 1/c$, $\text{exp}\{-cs^2 \} \geq 1 - cs^2$ and
\begin{align*}
 \int_{0}^{1/c} (s^2)^{c -1/2 - 1} \text{exp}\left\{-cs^2 \right\} ds^2 \geq& \int_{0}^{1/c} (s^2)^{c -1/2 - 1} \left(1 - cs^2 \right) ds^2 \\
 =& \int_{0}^{1/c} (s^2)^{c -1/2 - 1} - c\int_{0}^{1/c} (s^2)^{c -1/2}ds^2 \\
 =& \frac{\left(1/c\right)^{c - 1/2}}{c - 1/2} - \frac{1}{c - 1/2}\text{lim}_{a\rightarrow 0} a^{c - 1/2}- c\frac{\left(1/c\right)^{c - 1/2 + 1}}{c - 1/2 + 1}+ \\
 &\frac{c}{c - 1/2 + 1}\underbrace{\text{lim}_{a\rightarrow 0}a^{c - 1/2 + 1}}_{=0 \text{ for } c > 0} \\
  =& \left\{\begin{array}{cc}
  +\infty & 0 < c < 1/2 \\
  \frac{\left(1/c\right)^{c - 1/2}}{c - 1/2} -  \frac{1}{c - 1/2}\mathbbm{1}_{\left\{c = 1/2 \right\}} - \frac{\left(1/c\right)^{c - 1/2}}{c - 1/2 + 1} & c \geq 1/2
  \end{array}\right.
\end{align*}

\subsubsection{Proof of Proposition 2.2}

Now we can evaluate the marginal density $p(\bs \beta= \bs b |c, \bs \Omega)$ when $b_j = 0$ for some $j \in \{1, \dots, p\}$. Without loss of generality, set $\beta_1 = 0$. Letting  $\bs b_{-1}$ and $\bs s_{-1}$ refer to the vectors $\bs b$ and $\bs s$ each with the first element removed and $(\bs \Omega^{-1})_{-1,-1}$ be the matrix $\bs \Omega^{-1}$ with the first row and column removed. 
For $\bs b = (0,\bs b_{-1})$,
\footnotesize
\begin{align*}
p\left(\left(\beta_1 = 0, \bs b_{-1}\right) | c, \bs \Omega \right) &\propto \int \frac{\left(s^2_i\right)^{c - 1}}{\prod_{i = 1}^p s_i}\text{exp}\left\{-\frac{1}{2} \left(\bs b'\text{diag}\left\{\bs 1/\bs s\right\}\bs \Omega^{-1} \text{diag}\left\{\bs 1/\bs s \right\}\bs b\right) - \sum_{j = 1}^pcs^2_j \right\} d\bs s \\
= \int &\left(\prod_{j = 2}^p \left(s^2_j\right)^{c - 1/2 - 1} \right)\text{exp}\left\{-\frac{1}{2} \left(\bs b_{-1}'\text{diag}\left\{\bs 1/\bs s_{-1}\right\}\left(\bs \Omega^{-1}\right)_{-1,-1} \text{diag}\left\{\bs 1/\bs s_{-1}\right\}\bs b_{-1}\right) - \sum_{j = 2}^p cs^2_j \right\} \\
&\underbrace{\int_{0}^{\infty} \left(s^2_1\right)^{c - 1/2 - 1}\text{exp}\left\{-cs^2_1 \right\}ds_1}_{(*)} d\bs s_{-1}.
\end{align*}
\normalsize
Applying Lemma~\ref{lemma:infstng}, the term denoted by $(*)$ evaluates to $+\infty$ for every value of $\bs s_{-1}$.

\subsubsection{Proof of Proposition 2.4}

Proposition 2.4 follows from Proposition 2.2 and Lemma~\ref{lemma:infstng}. For $\bs \beta = (0, \bs \beta_{-1})$, 
\footnotesize
\begin{align*}
p\left( \beta_1 = 0 | c, \bs \Omega \right) &\propto \int \frac{\left(s^2_i\right)^{c - 1}}{\prod_{i = 1}^p s_i}\text{exp}\left\{-\frac{1}{2} \left(\bs \beta'\text{diag}\left\{\bs 1/\bs s\right\}\bs \Omega^{-1} \text{diag}\left\{\bs 1/\bs s \right\}\bs \beta\right) - \sum_{j = 1}^pcs^2_j \right\} d\bs s d \bs \beta_{-1} \\
= \int &\left(\prod_{j = 2}^p \left(s^2_j\right)^{c - 1/2 - 1} \right)\text{exp}\left\{-\frac{1}{2} \left(\bs \beta_{-1}'\text{diag}\left\{\bs 1/\bs s_{-1}\right\}\left(\bs \Omega^{-1}\right)_{-1,-1} \text{diag}\left\{\bs 1/\bs s_{-1}\right\}\bs \beta_{-1}\right) - \sum_{j = 2}^p cs^2_j \right\} \\
&\underbrace{\int_{0}^{\infty} \left(s^2_1\right)^{c - 1/2 - 1}\text{exp}\left\{-cs^2_1 \right\}ds_1}_{(*)} d\bs s_{-1}d \bs \beta_{-1}.
\end{align*}
\normalsize

Again, $(*)$ evaluates to $+\infty$ for any value of $\bs s_{-1}$ and does not depend at all on $\bs \beta_{-1}$.

\section{Kurtosis of SHP $\beta_j$}

\subsection{SNG $\beta_j$}

When $s^2_j \sim \text{gamma}(c, c)$, we have:
\begin{align*}
\int_0^\infty s_j^4 \frac{c^c}{ \Gamma\left(c\right)}\left(s^2_j\right)^{c - 1} \text{exp}\left\{-cs^2_j \right\}d s^2_j &= \frac{c^c}{ \Gamma\left(c\right)}\int_0^\infty \left(s^2_j\right)^{c + 2 - 1} \text{exp}\left\{-cs^2_j \right\}d s^2_j \\
&= \left(\frac{c^{c}\Gamma\left(c + 2\right)}{c^{c + 2} \Gamma\left(c\right)}\right) \\
&= c^{-2}\left(\Gamma\left(c + 2\right)/\Gamma\left(c\right)\right) \\
&= \left(c + 1\right)/c.
\end{align*}
A standard normal random variable has $\mathbb{E}[z^4_j] = 3$. It follows that

\begin{align*}
\mathbb{E}\left[\beta_j^4\right]/\mathbb{E}\left[\beta^2_j\right]^2 &= 3\left(c + 1\right)/c.
\end{align*}

\subsection{SPB $\beta_j$}

The kurtosis of a random variable is not a function of the overall scale, so without loss of generality, let $\mathbb{V}[\beta_j] = 1$.
When $\bs \beta$ is distributed according to an $SPB$ prior, elements $\beta_j$ each have a power/bridge distribution with density
\begin{align*}
p\left(\beta_j | q\right) = \left(\frac{q}{2}\right)\sqrt{\frac{\Gamma\left(3/q\right)}{\Gamma\left(1/q\right)^3}} \text{exp}\left\{- \left(\frac{\Gamma\left(3/q\right)}{\Gamma\left(1/q\right)} \right)^{q/2} \left|\beta_j\right|^q\right\}.
\end{align*}

It follows that 
\begin{align*}
\mathbb{E}\left[\beta_j^{2k}\right] =& \int_{-\infty}^{\infty} \left(\frac{q}{2}\right)\sqrt{\frac{\Gamma\left(3/q\right)}{\Gamma\left(1/q\right)^3}}  \beta_j^{2k}\text{exp}\left\{- \left(\frac{\Gamma\left(3/q\right)}{\Gamma\left(1/q\right)} \right)^{q/2} \left|\beta_j\right|^q\right\} d\beta_j \\
=& \int_{0}^{\infty}q\sqrt{\frac{\Gamma\left(3/q\right)}{\Gamma\left(1/q\right)^3}}  \beta_j^{2k}\text{exp}\left\{- \left(\frac{\Gamma\left(3/q\right)}{\Gamma\left(1/q\right)} \right)^{q/2} \beta_j^q\right\} d\beta_j \\
=& \int_{0}^{\infty} \sqrt{\frac{\Gamma\left(3/q\right)}{\Gamma\left(1/q\right)^3}}  \gamma_j^{\left(2k + 1\right)/q - 1}\text{exp}\left\{- \left(\frac{\Gamma\left(3/q\right)}{\Gamma\left(1/q\right)} \right)^{q/2} \gamma_j\right\} d\gamma_j\text{, }\gamma_j = \beta_j^q\text{, } \gamma^{1/q}_j = \beta_j \\
=& \sqrt{\frac{\Gamma\left(3/q\right)}{\Gamma\left(1/q\right)^3}} \left(\frac{\Gamma\left(\left(2k + 1\right)/q \right)}{\left(\frac{\Gamma\left(3/q\right)}{\Gamma\left(1/q\right)} \right)^{\left(2k + 1\right)/2}}\right) \\
=& \Gamma\left(3/q \right)^{\left(- 2k \right)/2} \Gamma\left(1/q\right)^{\left(2k -2\right)/2}\Gamma\left(\left(2k + 1\right)/q \right).
\end{align*}
It follows that the kurtosis of $\beta_j$ is 
\begin{align*}
\frac{\mathbb{E}\left[\beta_j^4\right]}{\mathbb{E}\left[\beta_j^2\right]^2} =& 
\frac{\Gamma\left(3/q \right)^{\left(- 4 \right)/2} \Gamma\left(1/q\right)^{\left(4 -2\right)/2}\Gamma\left(\left(4 + 1\right)/q \right)}{\Gamma\left(3/q \right)^{-2} \Gamma\left(1/q\right)^{2 -2}\Gamma\left(\left(2 + 1\right)/q \right)^2} \\
=& 
\frac{\Gamma\left(1/q\right)\Gamma\left(5/q \right)}{\Gamma\left(3/q \right)^2}.
\end{align*}

\section{Expectation of $s_j$}

\subsection{SNG $\beta_j$ with $\mathbb{V}[\beta_j] = 1$} 

When $s^2_j \sim \text{gamma}(c, c)$, we have:
\begin{align*}
\int_0^\infty s_j \frac{c^c}{ \Gamma\left(c\right)}\left(s^2_j\right)^{c - 1} \text{exp}\left\{-cs^2_j \right\}d s^2_j &= \frac{c^c}{ \Gamma\left(c\right)}\int_0^\infty \left(s^2_j\right)^{c + 1/2 - 1} \text{exp}\left\{-cs^2_j \right\}d s^2_j \\
&= \left(\frac{c^{c}\Gamma\left(c + 1/2\right)}{c^{c + 1/2} \Gamma\left(c\right)}\right) \\
&= c^{-1/2}\left(\Gamma\left(c + 1/2\right)/\Gamma\left(c\right)\right)
\end{align*}

\subsection{SPB $\beta_j$ with $\mathbb{V}[\beta_j] = 1$}

This is a little challenging because working with the stable distribution directly is difficult. However, given knowledge of normal moments and the marginal distribution of $\beta$, we can ``back out'' $\mathbb{E}[s]$. Let $\beta = s z$, where $z$ is a standard normal random variable and $1/s^2$ has an $\alpha$-stable distribution on the positive real line. When the stable prior is parametrized to yield $\mathbb{E}[\beta^2] = 1$, we have:
\begin{align*}
p\left(\beta | q\right) = \left(\frac{q}{2}\right)\sqrt{\frac{\Gamma\left(3/q\right)}{\Gamma\left(1/q\right)^3}}\text{exp}\left\{-\left(\frac{\Gamma\left(3/q\right)}{\Gamma\left(1/q\right)}\right)^{q/2}\left|\beta\right|^q\right\}.
\end{align*}

Now, to determine $\mathbb{E}[s]$, note that $\mathbb{E}[|\beta|] = \mathbb{E}[s]\mathbb{E}[|z|]$. We have
\begin{align*}
\mathbb{E}\left[\left|\beta\right| \right] &= 2\int_0^\infty \beta p\left(\beta | q\right)d\beta \\
&= q\sqrt{\frac{\Gamma\left(3/q\right)}{\Gamma\left(1/q\right)^3}}\int_0^\infty \beta \text{exp}\left\{-\left(\frac{\Gamma\left(3/q\right)}{\Gamma\left(1/q\right)}\right)^{q/2}\beta^q\right\}d\beta \\
&= \left(\sqrt{\frac{\Gamma\left(3/q\right)}{\Gamma\left(1/q\right)^3}}\right)\left(\frac{\Gamma\left(2/q\right)\Gamma\left(1/q\right)}{\Gamma\left(3/q\right)}\right) \\
&= \frac{\Gamma\left(2/q\right)}{\sqrt{\Gamma\left(1/q\right)\Gamma\left(3/q\right)}}. 
\end{align*}

According to \href{https://en.wikipedia.org/wiki/Normal_distribution}{Wikipedia}, $\mathbb{E}[|z|] = \sqrt{2/\pi}$. It follows that 
\begin{align*}
\mathbb{E}\left[s\right] = \sqrt{\frac{\pi}{2}}\left(\frac{\Gamma\left(2/q\right)}{\sqrt{\Gamma\left(1/q\right)\Gamma\left(3/q\right)}}\right).
\end{align*}

\section{Prior Conditional Distributions of $s^2_j$ and $s_j$}

\subsection{SPB $\beta_j$}

When using the SPB prior, we have $s^2_j \stackrel{d}{=}\Gamma(1/(2\alpha)) \xi_j^{\frac{1 - \alpha}{\alpha}}/(2\Gamma(3/(2\alpha)))$, where $\xi_j \sim \text{gamma}((1 + \alpha)/(2\alpha), f(\delta_j | \alpha) )$ and $p(\delta_j | \alpha) \propto f(\delta_j | \alpha)^{\frac{\alpha - 1}{2\alpha}}$ on $(0, \pi)$. We want to compute the prior conditional distribution $\pi(s^2_j | \delta_j, \alpha)$ for use in the elliptical slice sampler for $\bs s$. We have:
\begin{align*}
s^2_j \stackrel{d}{=}&\Gamma\left(1/\left(2\alpha\right)\right) \xi_j^{\frac{1 - \alpha}{\alpha}}/\left(2\Gamma\left(3/\left(2\alpha\right)\right)\right) \\
=&\left(\left(\Gamma\left(1/\left(2\alpha\right)\right)/\left(2\Gamma\left(3/\left(2\alpha\right)\right)\right)\right)^{\frac{\alpha}{1 - \alpha}} \xi_j\right)^{\frac{1 - \alpha}{\alpha}} \\
\stackrel{d}{=}&\tilde{\xi}_j^{\frac{1 - \alpha}{\alpha}}\text{, }\tilde{\xi}_j \sim \text{gamma}\left(\frac{1 + \alpha}{2\alpha} , \left(\frac{2\Gamma\left(\frac{3}{2\alpha}\right)}{\Gamma\left(\frac{1}{2\alpha}\right)}\right)^{\frac{\alpha}{1 - \alpha}}f\left(\delta_j | \alpha\right)\right) \\
\end{align*}
We can find the density of $s^2_j$ via change of variables. We have
\begin{align*}
\tilde{\xi}_j = \left(s^2_j\right)^{\frac{\alpha}{1 - \alpha}}\text{, }d\tilde{\xi}_j = \left(\frac{\alpha}{1 - \alpha}\right)\left(s^2_j\right)^{\frac{\alpha}{1 - \alpha} - 1}ds^2_j.
\end{align*}
Then the prior conditional distribution is
\begin{align*}
\pi\left(s^2_j | \delta_j, \alpha\right) \propto &\left(\left(s^2_j \right)^\frac{\alpha}{1 - \alpha}\right)^{\frac{1 +\alpha}{2\alpha} - 1}\left(s^2_j\right)^{\frac{\alpha}{1 - \alpha} - 1}\text{exp}\left\{- \left(\frac{2\Gamma\left(\frac{3}{2\alpha}\right)s^2_j}{\Gamma\left(\frac{1}{2\alpha}\right)}\right)^{\frac{\alpha}{1 - \alpha}}f\left(\delta_j | \alpha\right) \right\}  \\
=&\left(s^2_j \right)^{\frac{1 +\alpha}{2\left(1 - \alpha\right)}  - 1}\text{exp}\left\{- \left(\frac{2\Gamma\left(\frac{3}{2\alpha}\right)s^2_j}{\Gamma\left(\frac{1}{2\alpha}\right)}\right)^{\frac{\alpha}{1 - \alpha}}f\left(\delta_j | \alpha\right) \right\} . 
\end{align*}

In practice, we end up working with $s_j$, so once more 
\begin{align*}
\pi\left(s_j | \delta_j, \alpha\right) \propto &s_j^{\frac{1 +\alpha}{1 - \alpha}  - 1} \text{exp}\left\{- \left(\frac{2\Gamma\left(\frac{3}{2\alpha}\right)s^2_j}{\Gamma\left(\frac{1}{2\alpha}\right)}\right)^{\frac{\alpha}{1 - \alpha}}f\left(\delta_j | \alpha\right) \right\}. 
\end{align*}

\subsection{SNG Prior $\beta_j$}
Under the SNG prior, we have
\begin{align*}
\pi\left(s^2_j | c \right) \propto \left(s^2_j\right)^{c - 1}\text{exp}\left\{-s^2_j c\right\}.
\end{align*}
In practice, we end up working with $s_j$, which has the density
\begin{align*}
\pi\left(s_j | c \right) \propto \left(s_j\right)^{2c - 1}\text{exp}\left\{-s^2_j c\right\}.
\end{align*}

\section{Maximum correlation for bivariate SHP $\bs \beta$}

\subsection{SNG $\bs \beta$} 

Consider $\bs \Omega = (1 - \omega)\bs I_2 + \omega \bs 1_2 \bs 1_2'$. When $s^2_j \sim \text{gamma}(c, c)$, we have:
\begin{align*}
\bs \Sigma = \left(\begin{array}{cc} 
1 & \omega c^{-1}\left(\Gamma\left(c + 1/2\right)/\Gamma\left(c\right)\right)^2  \\
\omega c^{-1}\left(\Gamma\left(c + 1/2\right)/\Gamma\left(c\right)\right)^2 & 1 
\end{array}\right),
\end{align*}
where $|\omega|\leq 1$ by positive semidefiniteness of $\bs \Omega$. It follows that for fixed $c$, the largest possible value of $|\rho|$ is $(\Gamma(c + 1/2)/\Gamma(c))^2/c$.

\subsection{SPB $\bs \beta$}

Consider $\bs \Omega = (1 - \omega)\bs I_2 + \omega \bs 1_2 \bs 1_2'$. We have:
\begin{align*}
\bs \Sigma = \left(\begin{array}{cc} 
1 & \omega \left(\frac{\pi}{2}\right)\left(\frac{\Gamma\left(2/q\right)}{\sqrt{\Gamma\left(1/q\right)\Gamma\left(3/q\right)}}\right)^2  \\
\omega \left(\frac{\pi}{2}\right)\left(\frac{\Gamma\left(2/q\right)}{\sqrt{\Gamma\left(1/q\right)\Gamma\left(3/q\right)}}\right)^2 & 1 
\end{array}\right),
\end{align*}
where $|\omega|\leq 1$ by positive semidefiniteness of $\bs \Omega$. It follows that for fixed $q$, the largest possible value of $|\rho|$ is $(\frac{\pi}{2})(\frac{\Gamma(2/q)}{\sqrt{\Gamma(1/q)\Gamma(3/q)}})^2$.

\section{Posterior Simulation Details}

\subsection{Univarate Slice Sampling}\label{sec:unislice}

In several places in this paper, we use a univariate slice sampling algorithm described in \cite{Neal2003} to simulate values of a random variable $x \in [a, b]$ with density proportional to $\text{exp}\{g(x)\}$. Given a previous value $\tilde{x}$, we can simulate a new value $\tilde{x}'$ from a full conditional using univariate slice sampling  as follows:
\begin{enumerate}
	\item Draw $e \sim \text{exp}(1)$.
	\item Draw $d \sim \text{uniform}(a, b)$. 
	\begin{itemize}
		\item If $g(\tilde{x}) - e \leq g(d)$, set $\tilde{x}' = d$.
		\item If $g(\tilde{x}) - e > g(d)$:
		\begin{enumerate}
			\item If $d < \tilde{x}$, set $a = d$, else if $d \geq \tilde{x}$ set $b = d$.
			\item Return to 3.
		\end{enumerate}
	\end{itemize}
\end{enumerate}

\subsection{Simulation from Full Conditional Distribution for $\delta_j$}

When using the SPB prior, we have $s^2_j \stackrel{d}{=}\Gamma(1/(2\alpha)) \xi_j^{\frac{1 - \alpha}{\alpha}}/(2\Gamma(3/(2\alpha)))$, where $\xi_j \sim \text{gamma}((1 + \alpha)/(2\alpha), f(\delta_j | \alpha) )$ and $p(\delta_j | \alpha) \propto f(\delta_j | \alpha)^{\frac{\alpha - 1}{2\alpha}}$ on $(0, \pi)$. Suppose $s^2_j$ is fixed. Then $\xi_j = (2\Gamma(3/(2\alpha))s^2_j/\Gamma(1/(2\alpha)))^{\frac{\alpha}{1 - \alpha}}$. Then we can write the full conditional distribution of $\delta_j$ as
\begin{align*}
p\left(\delta_j | \xi_j, q \right) \propto& f\left(\delta_j | \alpha\right)^{\frac{1 + \alpha}{2\alpha}} \text{exp}\left\{ - f\left(\delta_j | \alpha\right)\xi_j \right\} f\left(\delta_j | \alpha\right)^{\frac{\alpha - 1}{2\alpha}}
 \\
 =&f\left(\delta_j | \alpha\right)^{\frac{\alpha^2 - 1}{4\alpha^2}} \text{exp}\left\{ - f\left(\delta_j | \alpha\right)\xi_j \right\}.
\end{align*}

Apply the univariate slice sampling algorithm described in Section~\ref{sec:unislice} using $g( x) = \text{log}(f(x | \alpha)) - f(x | \alpha)\xi_j $ and initial values $a = 0$ and $b = \pi$.

\subsection{Simulation from Full Conditional Distribution for $\rho$}

In the real data application, I consider the setting where $\bs B$ and $\bs S$ are $p_1 \times p_2$ matrices and $\bs \beta = \text{vec}(\bs B/\bs S)$ has covariance matrix $\mathbb{V}[\bs \beta/\bs s] = \bs \Omega_2 \otimes \bs \Omega_1$, where $\bs \Omega_1$ is depends on a single autoregressive parameter, $\rho$, with $\omega_{1,ij} = \rho^{|i - j|}$. The full conditional distribution of $\rho$ is:
\begin{align*}
p\left(\rho | -\right) \propto&	\left| \bs \Omega_1\right|^{-\frac{p_2}{2}}\text{exp}\left\{-\frac{1}{2}\text{tr}\left(\bs \Omega_1^{-1}\left(\bs B/\bs S\right) \bs \Omega_2^{-1}\left(\bs B/\bs S\right)\right) \right\} \\
=&\left(1 - \rho^2 \right)^{-\frac{\left(p_1 - 1\right)p_2}{2}}\text{exp}\left\{-\frac{1}{2}\text{tr}\left(\bs \Omega_1^{-1}\left(\bs B/\bs S\right) \bs \Omega_2^{-1}\left(\bs B/\bs S\right)\right) \right\}\pi\left(\rho \right),
	\end{align*}
	where elements of $\bs \Omega_1^{-1}$ are given by
	\begin{align*}
	\omega^{11}_1 &= \omega_1^{pp} = \left(1 - \rho^2\right)^{-1} \\	
	\omega_1^{jj} &=  \left(1 - \rho^2\right)^{-1} \left(1 + \rho^2\right)&& j \neq 1, j\neq p \\
	\omega^{jk}_1 &= -\rho \left(1 - \rho^2\right)^{-1} &&\left|j - k\right| = 1
	\end{align*}
	and $\pi(\rho)$ is the density of an assumed prior distribution for $\rho$. We assume a uniform prior on $(-1, 1)$, i.e. $\text{beta}(1, 1)$ prior for $(\rho + 1)/2$.
	
\subsection{Coordinate Descent for $\bs s$}


For any element of $\bs s$, the 
\begin{align*}
\kappa_1s^{-2}_j + \kappa_2 s^{-1}_j + \kappa_3 \text{log}\left(s^2_j\right) + \kappa_4 s_j + \kappa_5 s_j^{\kappa_6}\text{, }\kappa_6 \neq 1
\end{align*}

The optimal value of $s$ will satisfy
\begin{align*}
-2\kappa_1s^{-3}_j - \kappa_2 s^{-2}_j + 2\kappa_3s_j^{-1} + \kappa_4 + \kappa_5 \kappa_6 s_j^{\kappa_6 - 1} &= 0 \\
-2\kappa_1 - \kappa_2 s_j + 2\kappa_3s_j^2 + \kappa_4 s^3_j + \kappa_5 \kappa_6 s_j^{\kappa_6 + 2} &= 0
\end{align*}

When $\kappa_6$ is an integer (as is the case for the SNG and SPN priors) we can find all values of $s$ that satisfy this quickly and exactly using fast polynomial root finding functions from the \texttt{polynom} package for \texttt{R}.

When $\kappa_6$ is \emph{not} an integer, find values of $s_{F}$ and $s_{C}$ that maximize 
\begin{align*}
\kappa_1s^{-2}_j + \kappa_2 s^{-1}_j + \kappa_3 \text{log}\left(s^2_j\right) + \kappa_4 s_j + \kappa_5 s_j^{\text{floor}\left(\kappa_6\right)} \text{ and }\\
\kappa_1s^{-2}_j + \kappa_2 s^{-1}_j + \kappa_3 \text{log}\left(s^2_j\right) + \kappa_4 s_j + \kappa_5 s_j^{\text{ceil}\left(\kappa_6\right)}
\end{align*}
respectively. This gives us an interval for the solution $s$. This gives us a (hopefully small) interval which contains the maximizing $s$ for the original problem, and we can compute the maximizing $s$ for the original problem using bisection.


\begin{align*}
\kappa_1 &= -\frac{1}{2}\beta^2_j \Omega^{jj} \\
\kappa_2 &= -\frac{1}{2}\beta_j \sum_{j' \neq j} \frac{\Omega^{jj'}\beta_j}{s_{j'}} \\
\kappa_3 &= \left\{\begin{array}{ll}
-\frac{1}{2}  &\text{SPN} \\
c - 1 & \text{SNG} \\
\frac{1 + \alpha}{2\left(1 - \alpha\right)} - 1 & \text{SPB}
\end{array}\right. \\
\kappa_4 &= \left\{\begin{array}{ll}
-\frac{1}{2} \sum_{j' \neq j} \Psi^{jj'}s_{j'} &\text{SPN} \\
0 & \text{SNG} \\
0 & \text{SPB}
\end{array}\right. \\
\kappa_5 &= \left\{\begin{array}{ll}
-\frac{1}{2}\Psi^{jj} &\text{SPN} \\
 -c & \text{SNG} \\
-\left(\frac{2\Gamma\left(\frac{3}{2\alpha}\right)}{\Gamma\left(\frac{1}{2\alpha}\right)}\right)^{\frac{\alpha}{1 - \alpha}}f\left(\delta_j | \alpha\right) & \text{SPB}
\end{array}\right. \\
\kappa_6 &= \left\{\begin{array}{ll}
 2 &\text{SPN} \\
 2 & \text{SNG} \\
 \frac{2\alpha}{1 - \alpha} & \text{SPB}
\end{array}\right.
\end{align*}

\section{Hyperparameter Maximum Marginal Likelihood}

The marginal likelihood is given by:
\begin{align*}
\int \text{exp}\left\{-h\left(\bs y | \bs X, \bs s \circ \bs z, \bs \phi\right) \right\}\left(\frac{1}{\sqrt{2\pi \left|\bs \Omega\right|}} \text{exp}\left\{-\frac{1}{2}\bs z'\bs \Omega^{-1}\bs z \right\}\right) p\left(\bs s | \bs \theta\right)d \bs z d \bs s.
\end{align*}
Given an initial value $\bs \Omega^{(0)}$, Gibbs-within-EM estimates of $\bs \Omega$ can be obtained by iterating the following until $||\bs \Omega^{(i + 1)} - \bs \Omega^{(i)}||_F$ converges:
\begin{itemize}
	\item using MCMC to simulate $M$ draws $(\bs z^{(1)},s^{(1)}), \dots, (\bs z^{(M)},\bs s^{(M)})$ from the joint posterior distribution of $(\bs z, \bs s)$ given $\bs \Omega^{(i)}$, $\bs \phi$ and $\bs \theta$, set $\widehat{\mathbb{E}}[\bs z \bs z' | \bs \Omega^{(i)}, \bs X, \bs y, \bs \phi , \bs \theta ] = \frac{1}{M}\sum_{j = 1}^M \bs z^{(j)}(\bs z^{(j)})'$;
	\item set $\bs \Omega^{(i + 1)} = \text{argmin}_{\bs \Omega > 0}\text{log}(|\bs \Omega|) + \frac{1}{2}\text{tr}(\bs \Omega^{-1}\widehat{\mathbb{E}}[\bs z \bs z' | \bs \Omega^{(i)}, \bs X, \bs y, \bs \phi , \bs \theta ] )$.
\end{itemize}
In practice, we will rarely assume $\bs \Omega$  is unstructured,  as doing so requires estimating $p + p(p - 1)/2$ unknown parameters from a single observation of a $p\times 1$ vector. Instead, we will parametrize $\bs \Omega$ as a function of lower-dimensional parameters, e.g. as a function of a single autoregressive parameter $\rho$ or as a function of several separable covariance matrices $\bs \Omega = \bs \Omega_K \otimes \dots \otimes \bs \Omega_1$ which correspond to variation along different modes of the matrix or tensor of regression coefficients $\bs B$. However, the general approach to maximum marginal likelihood estimation of $\bs \Omega$ remains the same, with minimization over $\bs \Omega$ in the second step replaced by minimization over the lower-dimensional components.

\clearpage
\section{\color{black}Seconds per Posterior Draw}

\begin{figure}[ht]
\centering
\includegraphics{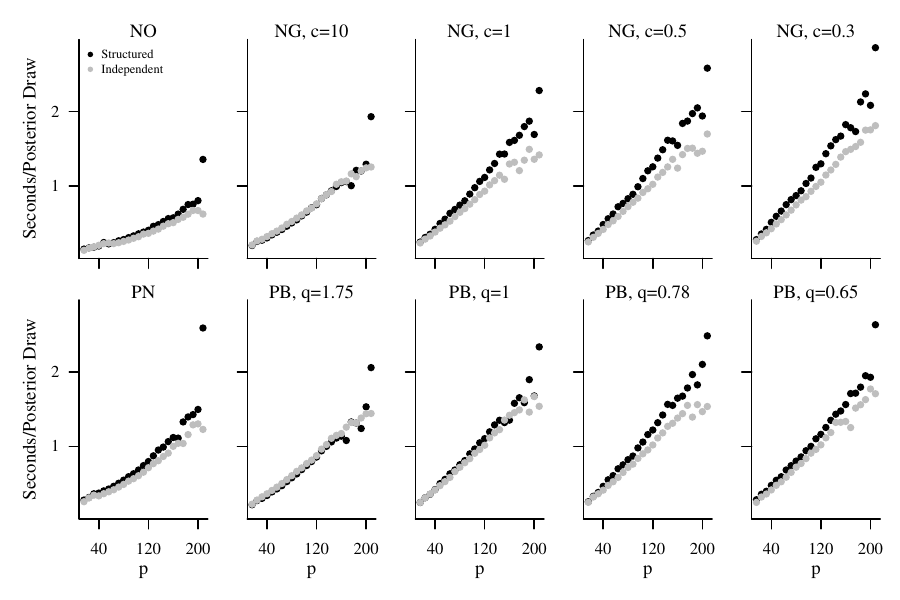}
\caption{\color{black}Seconds per posterior draw as a function of the total number of penalized covariates based on a single chain of $13,500$ simulated draws per prior setting, using the model and data described in Section 5 with the first $p_1 = 2,\dots, 26$ time points and all $p_2 = 8$ channels.}
\label{fig:timebyp}
\end{figure}

\clearpage
\section{\color{black}Minimum Effective Sample Sizes}

\begin{figure}[ht]
\centering
\includegraphics{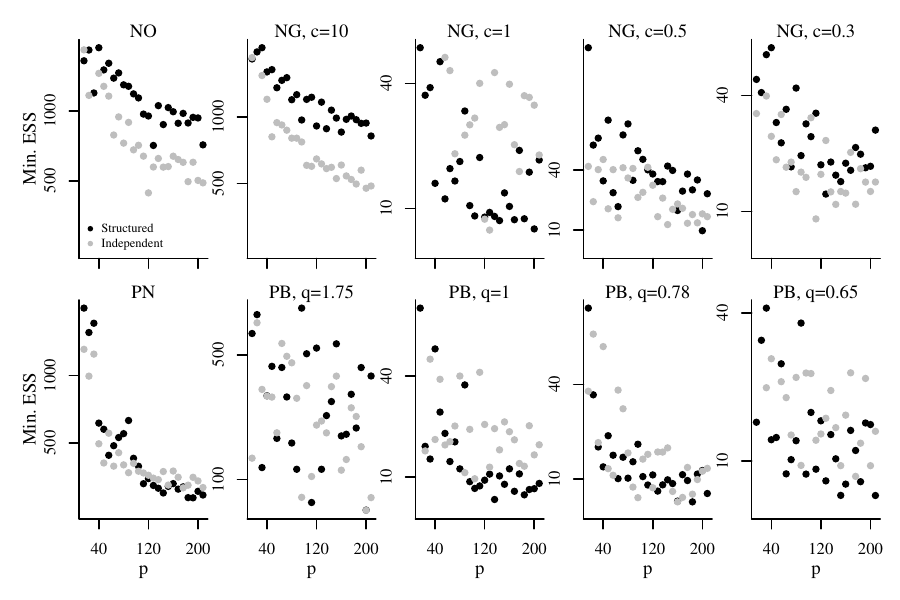}
\caption{\color{black}Minimum effective sample sizes as a function of the total number of penalized covariates based on a single chain of $13,500$ simulated draws per prior setting, using the model and data described in Section 5 with the first $p_1 = 2,\dots, 26$ time points and all $p_2 = 8$ channels.}
\label{fig:essbyp}
\end{figure}

\clearpage
\section{\color{black}Comparison to Independent Priors}

\begin{figure}[ht]
\centering
\includegraphics{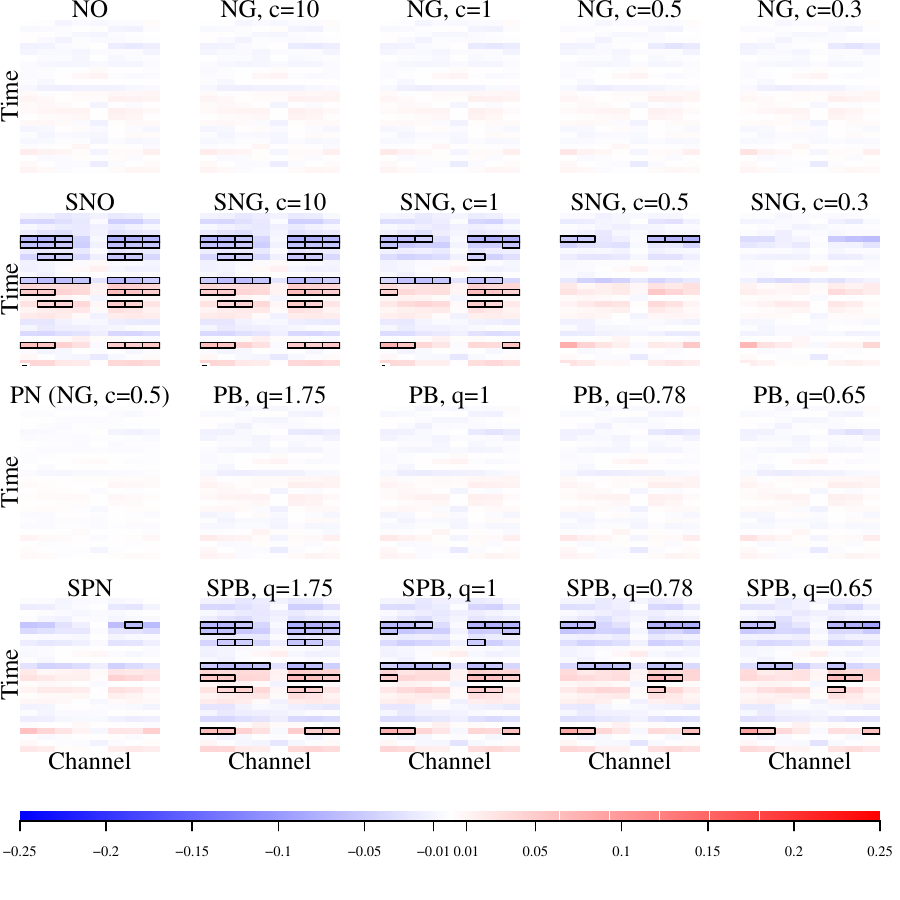}
\caption{\color{black}Approximate posterior means of $\bs B$. Boxes enclosing cells indicate approximate 90\% intervals that do not include zero.}
\label{fig:vmeansall}
\end{figure}

\section{\color{black}Effective Sample Sizes for Application}

\begin{table}[h!]
\footnotesize
\centering
\color{black}
\begin{tabular}{c|ccccc}
& \multirow{2}{*}{SNO} & \multicolumn{4}{c}{SNG} \\
&  & $c = 0.3$ & $c = 0.5$ & $c = 1$ & $c = 10$ \\ \hline \hline
$\boldsymbol \beta$ Only & $24,648.14$ & $976.34$ & $1,474.14$ & $1,980.6$ & $25,063.38$ \\
All Parameters & $18,975.88$ & $583.2$ & $797.4$ & $594.44$ & $19,371.21$ \\ \hline 
& \multirow{2}{*}{SPN} & \multicolumn{4}{c}{SPB}  \\
& & $q = 0.65$ & $q = 0.78$ & $q = 1$ & $q = 1.75$ \\ \hline \hline
$\boldsymbol \beta$ Only & $14,075.65$ & $1,300.39$ & $1,519.00$ & $2,342.22$ & $26,764.76$  \\
All Parameters & $1,839.08$ & $423.62$ & $329.33$ & $365.43$ & $218.41$\\ \hline
\end{tabular}
\caption{\color{black}Minimum effective sample sizes for 50,000 samples collected from 20 chains of 13,500 samples each, discarding initial $1,000$ samples from each chain as burn-in and thinning by a factor of $5$.}
\label{tab:ess}
\end{table} 

\end{appendix}

\end{document}